\documentclass[a4paper,11pt]{article}                                                      

   \usepackage[english]{babel}
\usepackage{amsfonts,amsmath,amsthm,graphicx,a4wide}                                                       

\usepackage{fancyhdr}
\usepackage{graphicx} 
\usepackage{amssymb}  
\usepackage{mathrsfs}
\usepackage[colorlinks=true]{hyperref}
\usepackage{dsfont} 
\usepackage{bbm}
\usepackage{xcolor}
\usepackage{hyperref}
\usepackage{enumerate}

\newtheorem{theorem}{Theorem}

\newtheorem{lemma}[theorem]{Lemma}
\newtheorem{proposition}[theorem]{Proposition} 

\newtheorem{definition}[theorem]{Definition}
\newtheorem{remark}[theorem]{Remark}
\newtheorem{example}[theorem]{Example}
\newcommand{\Tr}{\mathrm{Tr}}
\begin{document}

\title{Feedback exponential stabilization of GHZ states of multi-qubit systems}

\author{Weichao Liang, Nina H. Amini, and Paolo Mason
\thanks{W. Liang is with Laboratoire Analyse G\'eom\'etrie Mod\'elisation, CY Cergy Paris Universit\'e, 2, av. Adolphe Chauvin, 95302 Cergy-Pontoise, cedex, France.(e-mail: weichao.liang@u-cergy.fr).}
\thanks{N. H. Amini is with Laboratoire des signaux et syst\`emes, CNRS-CentraleSup\'elec-Universit\'e Paris-Sud, Universit\'e Paris-Saclay, 3, rue Joliot Curie, 91190 Gif-sur-Yvette, France (e-mail: nina.amini@l2s.centralesupelec.fr).}
\thanks{P. Mason is with Laboratoire des signaux et syst\`emes, CNRS-CentraleSup\'elec-Universit\'e Paris-Sud, Universit\'e Paris-Saclay, 3, rue Joliot Curie, 91190 Gif-sur-Yvette, France (e-mail: paolo.mason@l2s.centralesupelec.fr).}}

\maketitle

\begin{abstract}
In this paper, we consider stochastic master equations describing the evolution
of a multi-qubit system interacting with electromagnetic fields undergoing continuous-time measurements. By considering multiple $z$-type (Pauli $z$ matrix on different qubits) and $x$-type (Pauli $x$ matrix on all qubits) measurements and one control Hamiltonian, we provide general conditions on the feedback controller and the control Hamiltonian ensuring almost sure exponential convergence to a predetermined Greenberger-Horne-Zeilinger (GHZ) state, which is assumed to be  a common eigenstate of the measurement operators. We provide explicit expressions of feedback controllers satisfying such conditions. We also consider the case of only $z$-type measurements and multiple control Hamiltonians, and we discuss asymptotic convergence towards a predetermined GHZ state. Finally, we demonstrate the effectiveness of our methodology for a three-qubit system through numerical simulations.
\end{abstract}

\section{Introduction}
\label{sec:introduction}
Entanglement is an important property  with applications in e.g., quantum teleportation, quantum cryptography and quantum computation~\cite{nielsen2002quantum,bengtsson2017geometry}. For two-qubit systems, maximally entangled (correlated) states are the Bell states. For multi-qubit systems with $\mathbf{n}\geq 3$  qubits (i.e., $\mathbf{n}$ spin-$\frac12$ systems), such maximally entangled states are the Greenberger-Horne-Zeilinger (GHZ) states~\cite{greenberger1989going}. 

The evolution of an open quantum system undergoing indirect continuous-time measurements is described by quantum filtering theory~\cite{belavkin1989nondemolition,belavkin1983theory}. The mathematical tools to study open quantum systems are quantum stochastic calculus and quantum probability theory (developed by Hudson and Parthasarathy~\cite{hudson1984quantum}). Feedback stabilization of continuous-time open quantum systems has been studied in different papers~\cite{van2005feedback, armen2002adaptive, mirrahimi2007stabilizing, tsumura2008global,ahn2002continuous,yamamoto2007feedback,mabuchi2005principles,liang2018exponential,liang2019exponential}. 

Concerning stabilization of Bell states for two-qubit systems, a switching quantum feedback controller ensuring asymptotic stabilization towards the target state has been designed in~\cite{mirrahimi2007stabilizing}. In case of perfect measurements, in~\cite{yamamoto2007feedback} the authors constructed a continuous feedback controller to stabilize Bell states. With a more general approach, in~\cite{liang2019exponential_two-qubit}, we investigated the exponential and asymptotic stabilization towards a target Bell state. 

In this paper we derive some general conditions on the feedback controller and the control Hamiltonian enforcing the exponential convergence towards the target GHZ state. In addition, when only $z$-type measurements are available, we show a local stability result and we discuss asymptotic stabilization of the system to an arbitrary GHZ state. To our knowledge, this study provides the first results on exponential stabilization of GHZ states. 

The approach that we adopt here generalizes the method developed in~\cite{liang2019exponential} in presence of multiple channels and in the case in which the measurement operators possess degenerate eigenvalues. The degeneracy of the eigenvalues is due to the fact that the GHZ states are assumed to be common eigenstates of the measurement operators.
 In~\cite{liang2019exponential}, we studied feedback exponential stabilization of $N$-level quantum angular momentum systems towards an eigenstate of the measurement operator. Our approach is based on stochastic and geometric control tools. In contrast with previous methods, with this approach we can precise the rate of convergence, which is central in quantum information processing.  

This paper is organized as follows. In Section~\ref{SEC:SYS}, we introduce the stochastic model describing multi-qubit systems with multiple quantum channels in presence of imperfect measurements. In Section~\ref{Sec: Asymptotics without feedback}, we prove a quantum state reduction result with exponential rate in presence of both $z$-type and $x$-type measurements. In Section~\ref{sec:asgs}, we provide general conditions on the feedback controller and the control Hamiltonian ensuring the exponential stabilization of a predetermined GHZ state, we give the convergence rate, and explicit feedback controllers are obtained. In Section~\ref{SEC:1QC}, we consider the case where only $z$-type measurements are present and multiple control Hamiltonians act on the system. Simulations of a three-qubit system are provided in Section~\ref{SEC:SIM}.

\textit{Notations:}
The imaginary unit is denoted by $i$. We denote the  identity matrix by $\mathds{1}$. We denote the conjugate transpose of a matrix $A$ by $A^*.$ $A_{i,j}$ represents the element of the matrix $A$ at $i$-th row and $j$-th column. The function $\Tr(A)$ corresponds to the trace of a square matrix $A.$ The commutator of two square matrices $A$ and $B$ is denoted by $[A,B]:=AB-BA.$ For $x\in\mathbb{C}$, $\mathbf{Re}\{x\}$ is the real part of $x$ and $\mathbf{Im}\{x\}$ is the imaginary part of $x$. 

Given a $m\times n$ matrix $A$ and a  $p\times q$ matrix $B$, the Kronecker product $A\otimes B$ is defined as the block matrix whose $(i,j)$-th block is equal to $A_{ij}B$ (or, equivalently, $(A\otimes B)_{p(i-1)+r,q(j-1)+s} = A_{ij}B_{rs}$), for $i=1,\dots, m$ and $j=1,\dots, n$. The Kronecker product of $n$ copies of a matrix $A$ is indicated as $A^{\otimes n}$. Given $n$ matrices $A_1,\dots, A_n$ we use the notation $\bigotimes^{n}_{j=1}A_j = A_1\otimes \dots \otimes A_n$. We will sometimes use the Dirac notation $| \cdot \rangle$ to denote column vectors. The Pauli matrices are 
\begin{equation*}
\sigma_x = \left(\begin{array}{cc}0 &1\\ 1 &0 \end{array}\right),\ \sigma_y=\left(\begin{array}{cc}0 &-i\\ i &0 \end{array}\right),\ \sigma_z=\left(\begin{array}{cc}1 &0\\ 0 &-1 \end{array}\right).
\end{equation*}
We denote by $\mathrm{int}(\mathcal{S})$ the interior of a subset of a topological space and by $\partial \mathcal{S}$ its boundary.
\section{Multi-qubit systems with multiple quantum channels}
\label{SEC:SYS}
Here we consider a multi-qubit system undergoing continuous-time non-demolition measurements through $m$ quantum channels. The corresponding quantum state is described by the density matrix $\rho$, which belongs to the compact space $\mathcal{S}_N:=\{\rho\in\mathbb{C}^{N \times N}|\,\rho=\rho^*,\Tr(\rho)=1,\rho\geq0\}$ with $N=2^{\mathbf{n}}$ and $\mathbf{n}$ is the number of entangled qubits. The dynamics of the quantum state is described by the following matrix-valued stochastic differential equation, in It\^o form
\begin{equation}
d\rho_t=F_0(\rho_t)dt+\sum^m_{k=1}F_{k}(\rho_t)dt+\sum^m_{k=1}\sqrt{\eta_{k}}G_k(\rho_t)dW_{k}(t).
\label{SME}
\end{equation}
We suppose $W_t=(W_k(t))_{1\leq k\leq m}$ is a $m$-dimensional standard Wiener process and its components $W_k$ are mutually independent. The parameter $\eta_k\in(0,1]$ represents the measurement efficiency for the $k$-th channel.
The vector fields in \eqref{SME} are given by 
\begin{align*}
F_0(\rho)&:=-i[H_0,\rho]-i\sum^n_{j=1}u_j(\rho)[H_j,\rho],\\
 F_k(\rho)&:=L_k\rho {L_k}-L^2_k\rho/2-\rho L^2_k/2,\\
 G_k(\rho)&:=L_k\rho+\rho L_k-2\Tr(L_k\rho)\rho,
\end{align*}
where $H_0=H^*_0\in\mathbb{C}^{N \times N}$ denotes the free Hamiltonian, $u_j:=u_j(\rho_t)$ is the feedback controller taking values in $\mathbb{R}$, $H_k=H^*_k\in\mathbb{C}^{N \times N}$ with $k\in\{1,\dots,n\}$ are the control Hamiltonians, and $L_i=L^*_i\in\mathbb{C}^{N \times N}$ are the measurement operators associated with the $i$-th quantum channel.

If $u\in\mathcal{C}^1(\mathcal{S}_N,\mathbb{R}^n)$, the existence and uniqueness of the solution of~\eqref{SME} can be shown by similar arguments as in~\cite[Proposition 3.5]{mirrahimi2007stabilizing}. Moreover, it can be shown as in~\cite[Proposition 3.7]{mirrahimi2007stabilizing} that $\rho_t$ is a strong Markov process in $\mathcal{S}_N$.

In the following, we aim to stabilize the above system towards a target GHZ state. The GHZ states are a class of $N=2^{\mathbf{n}}$ entangled states described by the density matrices $\mathbf{GHZ}^{\pm}_{k}=\mathrm{ghz}^{\pm}_{k}\big(\mathrm{ghz}^{\pm}_{k}\big)^*$ for $k\in\{1,\dots,N/2\}$ where
\begin{equation*}
\mathrm{ghz}^{\pm}_{k}=\frac{1}{\sqrt{2}}\Big(\bigotimes^{\mathbf{n}}_{j=1}|k_j\rangle \pm  \bigotimes^{\mathbf{n}}_{j=1}|1-k_j\rangle \Big),
\end{equation*}
where $k_1=0$ and $k_2,\cdots,k_n$ represent the binary digits from left to right of the number $k-1.$ 

The family $\{\mathrm{ghz}^{\pm}_{k}\}_{1\leq k\leq N/2}$ is an orthogonal basis of $\mathbb{C}^{N}$. Throughout this paper we will make use of the following assumption.
\begin{enumerate}
\item[\textbf{(A0)}] $H_0,L_1,\dots,L_m$ are diagonal in $\{\mathrm{ghz}^{\pm}_{k}\}_{1\leq k\leq N/2}$.
\end{enumerate}
Before stating the notions of stability,
we recall that the Bures distance~\cite{bengtsson2017geometry} between two density matrices $\rho$ and $\sigma$ in $\mathcal{S}_N$ is defined as $d_B(\rho,\sigma) := (2-2\Tr( \sqrt{\sqrt{\rho}\sigma\sqrt{\rho}} ))^{1/2}$. This distance reduces to $
d_B(\rho,\sigma)= (2-2\sqrt{\Tr(\rho\sigma)}\,)^{1/2}$
when at least one among $\rho$ and $\sigma$ is a pure state. This Bures distance between $\rho$ and a set $E \subseteq \mathcal{S}_N$ is defined by 
$
d_B(\rho, E) := \min_{\sigma \in E} d_B(\rho,\sigma).
$
We denote the ball of radius $r>0$ around $E\subseteq \mathcal{S}_N$ as
$
B_r(E) := \{\rho \in \mathcal{S}_N|\, d_B(\rho,E) < r\}.$

Now we are ready to give some definitions of stochastic stability adapted from classical notions (see e.g.,~\cite{mao2007stochastic,khasminskii2011stochastic}) to our setting. 
\begin{definition}
Let $\bar{E}$ be an invariant set of System~\eqref{SME}, then $\bar{E}$ is said to be
\begin{enumerate}
\item[1.] 
\emph{locally stable in probability}, if for every $\varepsilon \in (0,1)$ and for every $r >0$, there exists  $\delta = \delta(\varepsilon,r)$ such that,
\begin{equation*}
\mathbb{P} \left( \rho_t \in B_r (\bar E) \textrm{ for } t \geq 0 \right) \geq 1-\varepsilon,
\end{equation*}
whenever $\rho_0 \in B_{\delta} (\bar E)$.

\item[2.]
\emph{almost surely asymptotically stable}, if it is locally stable in probability and,
\begin{equation*}
\mathbb{P} \left( \lim_{t\rightarrow\infty}d_B(\rho_t,\bar E)=0 \right) = 1,
\end{equation*}
whenever $\rho_0 \in \mathcal{S}_N$.

\item[3.]
\emph{exponentially stable in mean}, if for some positive constants $\alpha$ and $\beta$,
\begin{equation*}
\mathbb{E}(d_B(\rho_t,\bar E)) \leq \alpha \,d_B(\rho_0,\bar E)e^{-\beta t},
\end{equation*} 
whenever $\rho_0 \in \mathcal{S}_N$. The smallest value $-\beta$ for which the above inequality is satisfied is called the \emph{average Lyapunov exponent}. 

\item[4.]
\emph{almost surely exponentially stable}, if
\begin{equation*}
\limsup_{t \rightarrow \infty} \frac{1}{t} \log d_B(\rho_t,\bar E) < 0, \quad a.s.
\end{equation*}
whenever $\rho_0 \in \mathcal{S}_N$. The left-hand side of the above inequality is called the \emph{sample Lyapunov exponent} of the solution. 
\end{enumerate}
\end{definition}
Clearly any equilibrium  of~\eqref{SME}, that is any quantum state $\bar\rho$ satisfying $\sum^m_{k=0} F_k(\bar\rho)=0$ and $G_k(\bar\rho)=0$ for all $k\in\{1,\dots,m\}$, is a special case of invariant set.

In the following, we make use of Lyapunov techniques to study the stability properties of the system~\eqref{SME}. To analyze the variation of Lyapunov functions, a crucial tool is the infinitesimal generator. Given a stochastic differential equation $dq_t=f(q_t)dt+\sum_{k=1}^m g^k(q_t)dW_k(t)$,  where $W_k$ are independent Wiener processes and $q_t$ takes values in $Q\subset \mathbb{R}^p,$ the infinitesimal generator  is the operator $\mathscr{L}$ acting on twice continuously differentiable functions $V: Q \times \mathbb{R}_+ \rightarrow \mathbb{R}$ in the following way
\begin{equation*}
\begin{split}
\mathscr{L}V(q,t):=&\frac{\partial V(q,t)}{\partial t}+\sum_{i=1}^p\frac{\partial V(q,t)}{\partial q_i}f_i(q)\\
&+\frac12 \sum_{k=1}^m D^2 V(q,t) (g^k(q),g^k(q)),
\end{split}
\end{equation*}
where $D^2 V(q,t)$ is the Hessian of the function $V(\cdot,t)$. It\^o formula describes the variation of the function $V$ along solutions of the stochastic differential equation and is given as follows
$dV(q,t) = \mathscr{L}V(q,t)dt+\sum_{k=1}^m\nabla V(q,t)g^k(q)dW_k(t).$ From now on, the operator $\mathscr{L}$ is associated with Equation~\eqref{SME}.
 \section{Quantum state reduction}
\label{Sec: Asymptotics without feedback}

In this section, we discuss the stability properties of System~\eqref{SME} in the case in which the control action is turned off, i.e., when $u\equiv 0$.

We will make use of the \textit{variance functions} of the measurement operators $L_i$, defined as $\mathscr{V}_i(\rho):=\Tr(L^2_i\rho)-\Tr(L_i\rho)^2$.  By endowing the space of density matrices with the Hilbert-Schmidt norm and by a simple application of the Cauchy-Schwartz inequality it is easy to see that  $\mathscr{V}_i(\rho)=0$ if and only if $L_i\rho = \lambda\rho$, where $\lambda$ is an eigenvalue of $L_i$. The density matrices satisfying this condition are precisely those whose range is contained in the eigenspace of $L_i$ corresponding to $\lambda$.
Under Assumption~\textbf{(A0)}, a simple computation shows that
\begin{equation*}
\mathscr{L}\mathscr{V}_i(\rho)=-4\eta_i \mathscr{V}_i(\rho)^2-\sum_{k\neq i}\eta_k\Tr(L_iG_k(\rho))^2\leq -4\eta_i \mathscr{V}_i(\rho)^2,
\end{equation*}
which implies 
\begin{equation*}
\mathscr{L}\Big(\sum^m_{i=1}\mathscr{V}_i(\rho)\Big)\leq -4\sum^m_{i=1}\eta_i \mathscr{V}_i(\rho)^2\leq -4 \frac{\bar\eta}{m} \Big(\sum^m_{i=1}\mathscr{V}_i(\rho)\Big)^2,
\end{equation*} 
where $\bar \eta:=\min\{\eta_1,\dots,\eta_m\}>0$. By the stochastic LaSalle-type theorem~\cite{mao1999stochastic}, it follows that $\sum^m_{i=1}\mathscr{V}_i(\rho_t)$ converges to zero almost surely, when $t$ goes to infinity. In other words, $\rho_t$ converges almost surely to the set of density matrices whose range is contained in a common eigenspace of the operators $L_i$, for $i=1,\dots,m$.
Hence, in order to enforce the convergence towards the set of GHZ states, a natural hypothesis is that the common eigenspaces of the operators $L_i$ reduce to the one-dimensional spaces generated by the GHZ states.

In the following, we take $m=m_z+1$, $L_i:=\sqrt{M_i}L^{(i)}_z$ for $i=1,\dots,m_z$ and $L_m:=\sqrt{M_m}L_x$ with $M_i>0$ for $i=1\dots,m$ describing the strength of measurements. Here $L_x=\sigma_x^{\otimes \mathbf{n}}$ and each  $L^{(i)}_z$ belongs to the space generated by the $N$-dimensional diagonal matrices of the form $\bigotimes^{\mathbf{n}}_{j=1}\sigma^j$, where $\sigma^j\in\{\sigma_z,\mathds{1}\}$ and an even number of copies of $\sigma_z$ appear in the Kronecker product. These matrices form a set of $\sum^{\lfloor \mathbf{n}/2 \rfloor}_{j=0}{\mathbf{n}\choose 2j}=2^{\mathbf{n}-1}$ linearly independent elements spanning the space of matrices of the form $\mathrm{diag}\{x_1,\dots,x_{N/2},x_{N/2}\dots,x_1\}$ with $x_i\in\mathbb{R}$.  We denote 
\begin{equation}
\mathbf{L}_z:=\textstyle\sum^{m_z}_{i=1}L^{(i)}_z=
\mathrm{diag}(l_1,\dots,l_{N/2},l_{N/2},\dots,l_1),
\label{LzLx}
\end{equation} and we suppose the following 
\begin{enumerate}
\item[\textbf{(A1)}] 
 $l_i\neq l_j$ if $i\neq j$. \end{enumerate}
We denote $\ell>0$ the minimum difference, in absolute value, between any two different $l_i$ and $l_j$, i.e., $\ell:=\min_{i\neq j}\{|l_i-l_j|\}$. 

For $\mathbf{n}$-qubit systems, we denote the set of all GHZ states by 
$
\bar{E}_{\mathbf{n}}:=\{\mathbf{GHZ}^{\pm}_1,\dots,\mathbf{GHZ}^{\pm}_{N/2}\}.
$
Since, up to multiplicative constants, the GHZ states are the only common eigenstates of the measurement operators $L_i$ for $i=1,\dots,m$, we deduce from the above argument that the solution of the $\mathbf{n}$-qubit system converges to $\bar{E}_{\mathbf{n}}$ almost surely, when $t$ goes to infinity. Note that, if we only consider the z-type measurements  $L_i=\sqrt{M_i}L^{(i)}_{z}$ for $i=1,\dots,m_z$, then the system converges to the larger set given by the union of the sets $\mathcal{T}_k:=\{\rho\in\mathcal{S}_N|\,\Lambda_{k}(\rho)=1\}$, where 
\begin{equation}
\Lambda_{k}(\rho):=\rho_{k,k}+\rho_{\bar{k},\bar{k}} 
\label{eq:lambda}
\end{equation}
with $\bar{k}:=N+1-k$ and $k\in\{1,\dots,N/2\}$.
 
LaSalle theorem does not provide any information concerning the rate of convergence towards the set of equilibria. Hence, by following the alternative approach introduced  in~\cite[Theorem 5.1]{liang2019exponential}, in the rest of this section we will prove the almost sure exponential convergence towards $\bar{E}_{\mathbf{n}}$.

Firstly, we provide some invariance properties of the solution of System~\eqref{SME} via the following lemma. 
\begin{lemma}
Assume $u\equiv 0$ and let $k\in \{1,\dots,N/2\}.$ If $\Tr(\rho_0\mathbf{GHZ}^{\pm}_k)=0$ then $\mathbb{P}\big(\Tr(\rho_t\mathbf{GHZ}^{\pm}_k)=0, \forall\, t\geq 0 \big)=1.$ If, 
on the contrary, $\Tr(\rho_0\mathbf{GHZ}^{\pm}_k)>0$ then $\mathbb{P}\big(\Tr(\rho_t\mathbf{GHZ}^{\pm}_k)>0, \forall\, t\geq 0 \big)=1.$
\label{InvariantU=0Z}
\end{lemma}
\proof
For any $k\in\{1,\dots,N/2\}$ and $u\equiv0$, the dynamics of $\Tr(\rho_t\mathbf{GHZ}^{\pm}_k)$ is given by
\begin{equation*}
\begin{split}
&d\Tr(\rho_t\mathbf{GHZ}^{\pm}_k)\\
&=\,2\sum^{m_z}_{i=1}\sqrt{\eta_i M_i}\Tr(\rho_t\mathbf{GHZ}^{\pm}_k)(l^{(i)}_k-\Tr(L^{(i)}_z\rho_t))dW_{i}(t)\\
&~~~~+2\sqrt{\eta_m M_m}\Tr(\rho_t\mathbf{GHZ}^{\pm}_k)(\pm 1-\Tr(L_x\rho_t))dW_{m}(t),
\end{split}
\end{equation*}
where $L^{(i)}_z\mathbf{GHZ}^{\pm}_k=l^{(i)}_k\mathbf{GHZ}^{\pm}_k$ for $i\in\{1,\dots,m_z\}$ and $k\in\{1,\dots,N/2\}$.
In particular, we have 
\begin{align*}
&\left|2\sqrt{\eta_i M_i}\Tr(\rho_t\mathbf{GHZ}^{\pm}_k)(l^{(i)}_k-\Tr(L^{(i)}_z\rho_t))\right|\leq R\Tr(\rho_t\mathbf{GHZ}^{\pm}_k),\\
&\left|2\sqrt{\eta_2 M_2}\Tr(\rho_t\mathbf{GHZ}^{\pm}_k)(\pm 1-\Tr(L_x\rho_t))\right|\leq R\Tr(\rho_t\mathbf{GHZ}^{\pm}_k),
\end{align*}
for some $R>0$. This yields the first part of the lemma.

Given $\varepsilon>0$, consider any $\mathcal{C}^2$ function on $\mathcal{S}_N$ such that $V_{k}(\rho)=1/\Tr(\rho_t\mathbf{GHZ}^{\pm}_k)$ if $\Tr(\rho_t\mathbf{GHZ}^{\pm}_k)>\varepsilon$. A simple computation shows that $\mathscr{L}V_{k}(\rho)\leq KV_{k}(\rho)$ if $\Tr(\rho_t\mathbf{GHZ}^{\pm}_k)>\varepsilon$ for some positive constant $K$. By setting $f(\rho,t)=e^{-K t} V_{k}(\rho)$, one has $\mathscr{L}f\leq 0$ whenever $\Tr(\rho_t\mathbf{GHZ}^{\pm}_k)>\varepsilon$. From this fact, by applying the same arguments as in~\cite[Lemma 4.1]{liang2019exponential}, one proves that the probability of $\Tr(\rho_t\mathbf{GHZ}^{\pm}_k)$ becoming zero in a finite fixed time $T$ is proportional to $\varepsilon$ and, being the latter arbitrary, it must be zero. 
\hfill$\square$

We denote $V_x(\rho):=1-\Tr(L_x\rho)^2$ and recall that $\Lambda_{k}(\rho)=\rho_{k,k}+\rho_{\bar{k},\bar{k}}$ with $\bar{k}=N+1-k.$
Note that 
\begin{equation}
\Lambda_k(\rho)=\Tr((\mathbf{GHZ}^{+}_k+\mathbf{GHZ}^{-}_k)\rho)\geq0\label{eq1}
\end{equation}
and
\begin{equation}
\begin{split}
 V_x(\rho)=&\Tr((\mathds{1}-L_x)\rho)\Tr((\mathds{1}+L_x)\rho)\\
 =&4\sum^N_{k=1}\Tr(\mathbf{GHZ}^{-}_k\rho)\sum^N_{k=1}\Tr(\mathbf{GHZ}^{+}_k\rho)\geq0.
 \end{split}
\label{eq2}
\end{equation}
We now show the exponential convergence towards $\bar{E}_{\mathbf{n}}$ in mean and almost surely for $\mathbf{n}$-qubit systems when $u\equiv0$. 
\begin{theorem}[Exponential quantum state reduction]
For System~\eqref{SME} with $L_i=\sqrt{M_i}L^{(i)}_z$ for $i=1,\dots,m_z$ and $L_m=\sqrt{M_m}L_x$, $u\equiv0$ and $\rho_0 \in \mathcal{S}_N,$ the set $\bar{E}_{\mathbf{n}}$ is exponentially stable in mean and a.s. with average and sample Lyapunov exponent less than or equal to $-\min\{\Gamma_z \ell^2/2m_z\, ,\,2\eta_mM_m\}$ with $\Gamma_z:=\min_{i\in\{1,\dots,m_z\}}\{\eta_i M_i\}$. Moreover, the probability of convergence to $\bar{\boldsymbol\rho}\in\bar{E}_{\mathbf{n}}$ is $\Tr(\rho_0 \bar{\boldsymbol\rho})$.
\label{QSR}
\end{theorem}
\proof
Let $I:=\{k| \,\Lambda_k(\rho_0)=0 \}$ and $\mathcal  S_I:=\{\rho\in \mathcal S_N| \,\Lambda_k(\rho)=0\mbox{ if and only if } k\in I \}.$ By Lemma~\ref{InvariantU=0Z} and \eqref{eq1}, $\mathcal  S_I$ is a.s. invariant for System~\eqref{SME}. Similarly by the above lemma and \eqref{eq2}, the sets $\{\rho\in\mathcal{S}_N|\,V_x(\rho)=0\}$ and $\{\rho\in\mathcal{S}_N|\,V_x(\rho)>0\}$ are a.s. invariant for System~\eqref{SME}. Consider the candidate Lyapunov function
\begin{equation}
V(\rho):=\sum_{k\neq h}\sqrt{\Lambda_k(\rho)\Lambda_h(\rho)}+\sqrt{V_x(\rho)}.
\label{eq:lyapr}
\end{equation}
Note that $V(\rho)=0$ if and only if $\rho\in\bar{E}_{\mathbf{n}}$. Since $V(\rho)$ is twice continuously differentiable on the invariant sets $S_I \cap \{\rho\in\mathcal{S}_N|\, V_x(\rho)>0\}$ and $S_I \cap \{\rho\in\mathcal{S}_N|\,V_x(\rho)=0\}$, 
by a straightforward computation (see Appendix~\ref{App:ComputationQSR}),
we have
\begin{equation}
\mathscr{L}V(\rho)\leq -\bar{C}V(\rho),
\label{Eq:LV_QSR}
\end{equation}
where $\bar{C}:=\min\{\Gamma_z \ell^2/2m_z,2\eta_mM_m\}$. Thus, for all $\rho_0 \in \mathcal{S}_N$,  $\mathbb{E}\big(V(\rho_t)\big) = V(\rho_0)-\bar{C}\int^t_0 \mathbb{E}\big(V(\rho_s)\big)ds.$
In virtue of Gr\"onwall inequality, we have
$
\mathbb{E}\big(V(\rho_t)\big)\leq V(\rho_0) e^{-\bar{C}t}.
$
By a straightforward calculation, we can show that the candidate Lyapunov function is bounded from below and above by the Bures distance from $\bar{E}_{\mathbf{n}}$, more precisely, $C_1d_B(\rho,\bar{E}_{\mathbf{n}}) \leq V(\rho) \leq C_2d_B(\rho,\bar{E}_{\mathbf{n}}),$
where $C_1 = 1/8$ and $C_2= N(N/2-1)+4$. It implies $\mathbb{E}\big(d_B(\rho_t,\bar{E}_{\mathbf{n}})\big) \leq \frac{C_2}{C_1}d_B(\rho_0,\bar{E}_{\mathbf{n}})e^{-\bar{C}t}$ for all $\rho_0 \in \mathcal{S}_N,$ which means that the set $\bar{E}_{\mathbf{n}}$ is exponentially stable in mean with average Lyapunov exponent less than or equal to $-\bar{C}$. By employing the same arguments as in~\cite[Theorem 5.1]{liang2019exponential}, we obtain $\limsup_{t\rightarrow\infty}\frac{1}{t}\log d_B(\rho_t,\bar E_{\mathbf{n}}) \leq -\bar{C}$ almost surely, which means that the set $\bar E_{\mathbf{n}}$ is a.s. exponentially stable with sample Lyapunov exponent less than or equal to $-\bar{C}$. Finally, the probability of convergence to $\bar{\boldsymbol\rho}\in\bar{E}_{\mathbf{n}}$ is $\Tr(\rho_0 \bar{\boldsymbol\rho})$, this can be shown by the similar arguments as in the proof of~\cite[Theorem 5.1]{liang2019exponential}). The proof is complete.
\hfill$\square$
\section{Almost sure global exponential stabilization}~\label{sec:asgs}
Our aim here is to provide general conditions on the feedback controller and the control Hamiltonian to exponentially stabilize System~\eqref{SME} towards the target GHZ state $\bar{\boldsymbol\rho}\in\bar{E}_{\mathbf{n}}$. To this aim, firstly we establish some technical results concerning invariance  and reachability properties of the system. These results together with a local Lyapunov approach allow us to show the exponential convergence towards $\bar{\boldsymbol\rho}$ and to estimate the convergence rate under some assumptions on the feedback law. Finally, we design a parametrized family of feedback controllers satisfying such conditions for some choice of the control Hamiltonian. 

In the following, we take $n=1$, that is we assume that only one control input $u=u_1(\rho),$ together with a control Hamiltonian $H_1,$ is available. We consider the presence of $m=m_z+1$ quantum channels with $L_i=\sqrt{M_i}L^{(i)}_z$ for $i=1,\dots,m_z$ and $L_m=\sqrt{M_m}L_x$. In order to achieve our control goal, we need to ensure that the elements of $\bar{E}_{\mathbf{n}}\setminus\{\bar{\boldsymbol\rho}\}$ are no more equilibria  of the system, so that we make the following hypothesis on the feedback controller: 
\begin{enumerate}
\item[\textbf{(A2)}] 
$u\in\mathcal{C}^1(\mathcal{S}_N,\mathbb{R})$, $u(\bar{\boldsymbol\rho})=0$ and $u(\boldsymbol\rho)[H_1,{\boldsymbol\rho}]\neq0$ for all $\boldsymbol\rho\in \bar{E}_{\mathbf{n}}\setminus\{\bar{\boldsymbol\rho}\}$.
\end{enumerate}

\subsection{Preliminary results on invariance and non-invariance properties}
The following lemmas are analogous to the results in~\cite[Section~4]{liang2019exponential} and they concern invariance properties of System~\eqref{SME} involving the boundary 
$\partial\mathcal{S}_N:=\{\rho\in\mathcal{S}_N|\,\det({\rho})=0\}.$ Since their proofs are based on the same arguments, we omit them. 
\begin{lemma}
The set of positive-definite matrices is a.s. invariant for~\eqref{SME}. More in general, the rank of $\rho_t$ is a.s. non-decreasing.
\label{Lem:RankNonDec}
\end{lemma}
\begin{lemma}
Assume $u\in\mathcal{C}^1(\mathcal{S}_N,\mathbb{R})$. If $\eta_i=1$ for all $i\in\{1,\dots,m\}$, then $\partial\mathcal{S}_N$ is a.s. invariant for~\eqref{SME}.
\label{Lem:BoundaryInv}
\end{lemma}
\begin{lemma}
Assume that \textbf{(A2)} is satisfied. If the initial state satisfies $\rho_0 \neq \bar{\boldsymbol\rho},$ then
$
\mathbb{P}( \rho_t \neq \bar{\boldsymbol\rho}, \forall\, t\geq 0 )=1.
$
\label{Never reach with feedback}
\end{lemma}
The following lemma shows that pure states become immediately mixed states for the case $\eta_i\in(0,1)$ for all $i\in\{1,\dots,m\}$.
\begin{lemma}
Suppose that $\eta_i\in(0,1)$ for all $i\in\{1,\dots,m\}$ and \textbf{(A2)} is satisfied. Then for all $\rho_0\in\{\rho\in\mathcal{S}_N|\Tr(\rho^2)=1\}\setminus\{\bar{\boldsymbol\rho}\}$, the density matrix $\rho_t$ is mixed (i.e., $\Tr(\rho_t^2)<1$) for all $t>0$ almost surely.
\label{Exits pure state}
\end{lemma}
\proof
For all $\rho\in\{\rho\in\mathcal{S}_N|\Tr(\rho^2)=1\}$, there exists a vector $\xi\in\mathbb{C}^N$ such that $\rho=\xi\xi^*$. Then we have $\Tr(L_i\rho L_i\rho)= (\xi^*L_i\xi)^2=\Tr(L_i\rho)^2$ for $i=1,\dots,m$.
Thus, by It\^o formula, for all $\rho_0\in\{\rho\in\mathcal{S}_N|\Tr(\rho^2)=1\}\setminus\{\bar{\boldsymbol\rho}\}$, 
$
d\Tr(\rho^2_t)=-2\sum^{m}_{i=1}(1-\eta_i)\mathscr{V}_i(\rho_t)dt
$
where $\mathscr{V}_i(\rho):=\Tr(L_i^2\rho)-\Tr(L_i\rho)^2$. Since $\eta_i\in(0,1)$, we have 
$
-2\sum^{m}_{i=1}(1-\eta_i)\mathscr{V}_i(\rho)\leq 0
$
and the equality holds if and only if $\rho\in\bar{E}_{\mathbf{n}}$. By Lemma~\ref{Never reach with feedback}, Lemma~\ref{Lem:RankNonDec} and the fact that $\rho_t$ exits $\bar{E}_{\mathbf{n}}\setminus\{\bar{\boldsymbol\rho}\}$ immediately almost surely by \textbf {(A2)}, the density matrix $\rho_t$ becomes mixed immediately and remains mixed afterwards almost surely. The proof is complete.\hfill$\square$ 

\subsection{Reachability of the target state}
In this section, our aim is to show that the stochastic trajectories $\rho,$ solutions of Equation~\eqref{SME} enter any arbitrary neighborhoods of the target state $\boldsymbol{\bar\rho}$ with non-zero probability in finite time. Our method is based on the use of the celebrated support theorem~\cite{stroock1972support} which  interprets the stochastic solutions as limits of solutions of a  deterministic equation. 

Before stating the support theorem, we recall that any stochastic differential equation in It\^o form in $\mathbb R^K$
\begin{equation*}
dx_t=\widehat X_0(x_t)dt+\sum^n_{k=1}\widehat X_k(x_t)dW_k(t), \quad x_0 = x,
\end{equation*}
can be written in the following Stratonovich form~\cite{rogers2000diffusions2}
\begin{equation*}
dx_t = X_0(x_t)dt+\sum^n_{k=1}X_k(x_t) \circ dW_k(t), \quad x_0 = x,
\end{equation*}
where 
$X_0(x)=\widehat X_0(x)-\frac{1}{2}\sum^K_{l=1}\sum^n_{k=1}\frac{\partial \widehat X_k}{\partial x_l}(x)(\widehat X_k)_l(x)$, $(\widehat X_k)_l$ denoting the component $l$ of the vector $\widehat X_k,$ and $X_k(x)=\widehat X_k(x)$ for $k\neq 0$.

\medskip

\begin{theorem}[Support theorem~\cite{stroock1972support}]
Let $X_0(x)$ be a bounded and uniformly Lipschitz continuous function, and $X_k(x)$  be  twice continuously differentiable, with bounded derivatives, for $k\neq 0.$ Consider the Stratonovich equation
\begin{equation*}
dx_t = X_0(x_t)dt+\sum^n_{k=1}X_k(x_t) \circ dW_k(t), \quad x_0 = x.
\end{equation*}
Let $\mathbb{P}_x$ be the probability law of the solution $x_t$ starting at $x$. Consider in addition the associated deterministic control system
\begin{equation*}
\frac{d}{dt}x_{v}(t) = X_0(x_{v}(t))+\sum^n_{k=1}X_k(x_{v}(t))v_k(t), \quad x_v(0) = x.
\end{equation*}
with $v_k \in \mathcal{V}$, where $\mathcal{V}$ is the set of all locally bounded measurable functions  from $\mathbb{R}_+$ to $\mathbb{R}$. Now we define $\mathcal{W}_x$ as the set of all continuous paths from $\mathbb{R}_+$ to $\mathbb R^K$ starting at $x$, equipped with the topology of uniform convergence on compact sets, and $\mathcal{I}_x$ as the smallest closed subset of $\mathcal{W}_x$ such that $\mathbb{P}_x(x_{\cdot} \in \mathcal{I}_x)=1$. Then,
$
\mathcal{I}_x = \overline{ \{ x_{v}(\cdot)\in\mathcal{W}_x|\, v \in \mathcal{V}^n\} } \subset \mathcal{W}_x.
$
\label{Support thm}
\end{theorem}

Based on the above theorem, we consider the following deterministic equation
\begin{equation}
\dot{\rho}_{v}(t)=F_0(\rho_{v}(t))+\sum^m_{k=1}\widehat{F}_k(\rho_{v}(t))+\sum^m_{k=1}\sqrt{\eta_k}G_k(\rho_{v}(t))v_k(t),
\label{ODE}
\end{equation}
for $\rho_{v}(0)=\rho_0$, with $v_k(t)\in\mathcal{V}$ for $k=1,\dots,m,$ 
where 
\begin{equation*}
\begin{split}
\widehat{F}_k(\rho):=\,&(1-\eta_k)L_k\rho L_k\\
&-\frac{1+\eta_k}{2}(L^2_k\rho+\rho L^2_k)+2\eta_k \Tr(L^2_k\rho)\rho\\
&+2\eta_k\Tr(L_k \rho)(L_k\rho+\rho L_k-2\Tr(L_k\rho)\rho),\\
\end{split}
\end{equation*}
$F_0$ and $G_k$ are defined as in~\eqref{SME}. By Theorem~\ref{Support thm}, the set $\mathcal S_N$ is invariant for Equation~\eqref{ODE}.


\medskip

For $l\in\mathbb{Z}$ and $\xi\in\mathbb R^N,$ define 
\begin{equation*}
\begin{split}
\mathbf{M}_{l,\xi}\!:=\![&\xi, H_1\xi, L^{(1)}_zH_1\xi, \dots, L^{(m_z)}_zH_1\xi ,L_xH_1\xi , \\
&\dots,H_1^l\xi, L^{(1)}_zH_1^l\xi,\dots, L^{(m_z)}_zH_1^l\xi, L_xH_1^l\xi]
\end{split}
\end{equation*}
and recall that $L^{(i)}_z\mathbf{GHZ}^{\pm}_k=l^{(i)}_k\mathbf{GHZ}^{\pm}_k.$
\begin{lemma}
Let $\bar{\boldsymbol\rho}=\mathbf{GHZ}_{\mathbf k}^{\boldsymbol\epsilon}$ with $\boldsymbol\epsilon\in\{+,-\}$ and $\mathbf k\in\{1,\dots,N/2\}$ and set $\mathbf{P}^{(i)}_z:=\{\rho\in\mathcal{S}_N|\,l^{(i)}_{\mathbf k}-\Tr(L^{(i)}_z \rho)=0\}$ for $i=1,\dots,m_z$ and $\mathbf{P}_x:=\{\rho\in\mathcal{S}_N|\,\boldsymbol\epsilon 1-\Tr(L_x \rho)=0\}$. Assume that \textbf{(A1)} and \textbf{(A2)} are satisfied. Suppose that
\begin{enumerate}
\item[(i)] there exists at least one $k\in\{1,\dots,m\}$ such that $\nabla u\cdot G_k(\rho_0)\neq 0$, or $\nabla u\cdot \big( F_0(\rho_0)+\sum^m_{k=1}\widehat F_k(\rho_0)\big)\neq 0$ for any $\rho_0\in\{\rho\in\mathcal S_N| \, \Tr(\rho\bar{\boldsymbol\rho})=0, \, u=0\},$
\item[(ii)] there exists a finite $l\in\mathbb{Z}$ such that $\mathrm{rank}(\mathbf{M}_{l,\mathrm{ghz}_{\bf k}^{\boldsymbol \epsilon}})\geq N-1,$  
\item[(iii)] for all $\rho\in(\bigcap^{m_z}_{i=1}\mathbf{P}^{(i)}_z\cap\mathbf{P}_x)\setminus \{\bar{\boldsymbol\rho}\}$, the feedback controller and the control Hamiltonian satisfy
\begin{equation}
 2\Tr(\rho\bar{\boldsymbol\rho})\sum^{m_z}_{i=1}\eta_i\mathscr V_i(\rho)>u\Tr(i[H_1,\rho]\bar{\boldsymbol\rho}),
\label{Condition u_t}
\end{equation}
where $\mathscr{V}_i(\rho)=\Tr(L_i^2\rho)-\Tr(L_i\rho)^2.$
\end{enumerate}
Then, for all $r>0$ and any given initial state $\rho_0 \in \mathcal{S}_N,$
$
\mathbb{P}(\tau_{r} < \infty)=1,
$
where $\tau_{r}: = \inf\{t \geq 0|\, \rho_t \in B_r(\bar{\boldsymbol\rho}) \}$ and $\rho_t$ corresponds to the solution of System~\eqref{SME}.
\label{Passage lemma_2QC}
\end{lemma}
\proof
The proof is based on similar arguments as in~\cite[Lemma 6.1]{liang2019exponential}.  For $\rho_{0} \in B_r(\bar{\boldsymbol\rho})$, we have $\tau_{r} = 0$. Let us thus suppose that $\rho_{0} \in \mathcal{S}_N \setminus B_r(\bar{\boldsymbol\rho})$. We show that there exists $T\in (0,\infty)$ and $\zeta\in (0,1)$ such that $\mathbb{P}_{\rho_0}( \tau_{r}<T )>\zeta$. Let $\rho_v$ be a solution of Equation~\eqref{ODE} and consider the following equation 
\begin{equation}
\begin{split}
\Tr(\dot{\rho}_v(t)\bar{\boldsymbol\rho})=&-\Theta_u(\rho_v(t))+\sum^m_{k=1}\Theta_k(\rho_v(t))\\
&+\Tr(\rho_v(t)\bar{\boldsymbol\rho})\sum^m_{k=1}\sqrt{\eta_k M_k}P_k(\rho_v(t))v_k(t)
\end{split}
\label{deterministic X(rho)}
\end{equation}
where $v_k(t)$ belonging to $\mathcal{V}$ are the control inputs, $\Theta_u(\rho):=u\Tr(i[H_1,\rho]\bar{\boldsymbol\rho})$ and for $i\in\{1,\dots,m_z\},$ 
\begin{align*}
P_i(\rho)&:=l^{(i)}_{\mathbf k}-\Tr(L^{(i)}_z\rho), \\
P_m(\rho)&:=\boldsymbol\epsilon 1-\Tr(L_x\rho),\\
\Theta_i(\rho)&:=2\eta_i M_i\Tr(\rho\bar{\boldsymbol\rho})\big(\Tr({L^{(i)}_z}^2\rho)-{l^{(i)}_{\mathbf k}}^2+2\Tr(L^{(i)}_z\rho)P_i(\rho) \big),\\
\Theta_m(\rho)&:=4\eta_m M_m\Tr(\rho\bar{\boldsymbol\rho})\Tr(L_x\rho)P_m(\rho).
\end{align*} 
Consider the special case in which $\Tr(\rho_0\bar{\boldsymbol\rho})=0$. The technical conditions $(i),$ $(ii)$ and Lemma~\ref{Exits pure state} allow us to apply similar arguments as in the proof of~\cite[Proposition 4.5]{liang2019exponential}, in order to show the existence of a control input $v\in\mathcal{V}^m$ such that $\Tr(\rho_{v}(t)\bar{\boldsymbol\rho})>0$ for some $t>0$ arbitrarily small.  

First, by Lemma~\ref{Exits pure state} and the support theorem, we can suppose without loss of generality  that $\mathrm{rank}(\rho_v(t))\geq 2$ fot $t>0$ arbitrarily small. Moreover, by the  condition $(i),$ we can assume that $u(\rho_v(t))\neq 0.$ Next we proceed by contradiction, that is we show that $\Tr(\rho_{v}(t)\bar{\boldsymbol\rho})\equiv 0$ on an interval $[0,\delta]$ implies $\mathrm{rank}(\mathbf{M}_{l,\mathrm{ghz}_{\bf k}^{\boldsymbol \epsilon}})< N-1.$ Note that $\Tr(\rho_{v}(t)\bar{\boldsymbol\rho})= 0$ is equivalent to $\rho_v(t)\mathrm{ghz}_{\bf k}^{\boldsymbol \epsilon}= 0.$

Let $\xi\in\mathbb C^N$ such that $\rho_v(t)\xi\equiv 0$ on $[0,\delta].$  We next show that $\rho_v(t)L_i\xi\equiv 0$ for $i=1,\dots,m$ and $\rho_v(t)H_1\xi\equiv 0.$ Indeed, 
by simple calculations, one has $$\xi^*\dot{\rho}_v(t)\xi=\sum_{i=1}^{m}(1-\eta_i)\xi^*L_i\rho_{v}(t)L_i\xi=0,$$ which implies $\rho_{v}(t)L_i\xi=0.$ Moreover, as a consequence one also has
$$\dot{\rho}_v(t)\xi=i\rho_{v}(t)H_0\xi+iu(\rho_v(t))\rho_{v}(t)H_1\xi=0.$$ Let $\bar L$ be a linear combination of the matrices $L_i$ only admitting simple eigenvalues (its existence is guaranteed by the condition \textbf {(A1)}). Since the minimal polynomial of $\bar L$ has degree $N,$ one easily deduces that the matrices $\bar L^k$ for $k=0,\dots,N-1$ span the space of matrices diagonal with respect to the basis of GHZ states. Hence $\rho_{v}(t)H_0\xi=0$ and as $u(\rho_v(t))\neq 0,$ we have $\rho_v(t)H_1\xi=0.$ This yields a contradiction with the condition $(ii).$ Thus, without loss of generality, we suppose $\Tr(\rho_0\bar{\boldsymbol\rho})>0$.

Next we show that there exist a control input $v\in\mathcal{V}^{m}$ and a time $T\in(0,\infty)$ such that $\rho_{v}(t)\in B_r(\bar{\boldsymbol\rho})$ for $t\leq T$ in the two following separate cases.
\begin{enumerate}
\item Let $l_{\mathbf{k}}>l_k$ or $l_{\mathbf{k}}<l_k$ for all $k\neq \mathbf{k}$. Obviously, for all $\rho\in\bigcap^{m_z}_{i=1}\mathbf{P}^{(i)}_z$, we have $\sum^{m_z}_{i=1}P_{i}(\rho)=l_{\mathbf{k}}-\Tr(\mathbf{L}_z\rho)=0$. In this case, we have that $\bigcap^{m_z}_{i=1}\mathbf{P}^{(i)}_z\cap\mathbf{P}_x=\bar{\boldsymbol\rho}$.
Due to the compactness of $\mathcal{S}_N\setminus B_r(\bar{\boldsymbol\rho})$, $-\Theta_u(\rho)+\sum^m_{k=1}\Theta_k(\rho)$ is bounded from above in $\mathcal{S}_N\setminus B_r(\bar{\boldsymbol\rho})$ and at least one among $|P_i(\rho)|$ for $i=1,\dots,m$ is bounded from below. 
By choosing $v_i=KP_i(\rho)/\Tr(\rho\bar{\boldsymbol\rho})$ with $K>0$ sufficiently large and $v_j=0$ for $j\neq i$, we can guarantee that $\rho_{v}(t)\in B_r(\bar{\boldsymbol\rho})$ for $t\leq T$ with $T<\infty$ if $\Tr(\rho_0\bar{\boldsymbol\rho})>0$.

\item Now consider the general case. By~\eqref{Condition u_t} and the compactness of $\mathbf{C}_r:=(\bigcap^{m_z}_{i=1}\mathbf{P}^{(i)}_z\cap\mathbf{P}_x) \setminus B_r(\bar{\boldsymbol\rho})$, we obtain
\begin{align*}
\mathbf{m}:&=\min_{\rho\in\mathbf{C}_r}\Big(-\Theta_u(\rho)+\textstyle\sum^m_{k=1}\Theta_k(\rho)\Big)\\
&=\min_{\rho\in\mathbf{C}_r}\Big(\textstyle2\sum^{m_z}_{i=1}\eta_i\mathscr V_i(\rho)\Tr(\rho\bar{\boldsymbol\rho})-u\Tr(i[H_1,\rho]\bar{\boldsymbol\rho})\Big)\\
&>0.
\end{align*}
Then we define an open set $\mathbf{U}\subseteq\mathcal{S}_N$ containing $\mathbf{C}_r$,
\begin{equation*}
\mathbf{C}_r\subseteq\mathbf{U}:=\left\lbrace\rho\in\mathcal{S}_N\left|\,-\Theta_u(\rho)+\textstyle \sum^m_{k=1}\Theta_k(\rho)>\mathbf{m}/2\right.\right\rbrace.
\end{equation*} 
Thus, setting $v_1(t)=v_2(t)=\dots=v_m(t)=0$ whenever $\rho_{v}(t)\in\mathbf{U}$, we have
\begin{equation*}
\Tr(\dot{\rho}_v(t)\bar{\boldsymbol\rho})=-\Theta_u(\rho_{v}(t))+\sum^m_{k=1}\Theta_k(\rho_{v}(t))>\frac{\mathbf{m}}{2}.
\end{equation*}
Moreover, $(\mathcal{S}_N\setminus B_r(\bar{\boldsymbol\rho}))\setminus\mathbf{U}$ is compact, then $-\Theta_u(\rho)+\textstyle \sum^m_{k=1}\Theta_k(\rho)$ is bounded from above and at least one among $|P_i(\rho)|$ for $i=1,\dots,m$ is bounded from below in this domain. Similarly to $1),$ we can take  $v_i=KP_i(\rho)/\Tr(\rho\bar{\boldsymbol\rho})$ with $K>0$ sufficiently large and $v_j=0$ for $j\neq i.$ The proposed inputs  guarantee that $\rho_{v}(t)\in B_r(\bar{\boldsymbol\rho})$ for $t\leq T$ with $T<\infty$ if $\Tr(\rho\bar{\boldsymbol\rho})>0$.

\end{enumerate}
Summing up, we have shown the existence of some  $T>0$ such that for all $\rho_{0}\in \mathcal S_N\setminus B_r(\bar{\boldsymbol\rho})$, there exists $v\in\mathcal{V}^m$ steering the system from $\rho_0$ to $B_r(\bar{\boldsymbol\rho})$ by time $T.$  Finally, the arguments provided in~\cite[Lemma 6.1]{liang2019exponential} allow to show that $
\mathbb{P}_{\rho_0}( \tau_{r}<\infty )=1.
$
\hfill$\square$
\subsection{Main results on feedback exponential stabilization}

By combining the previous lemmas and following the same arguments as in~\cite[Theorem 6.2]{liang2019exponential}, we get the following general result concerning the exponential stabilization towards the GHZ state $\bar{\boldsymbol\rho}$. 
\begin{theorem}
Assume that $\rho_0\in \mathcal S_N$ and the assumptions of Lemma~\ref{Passage lemma_2QC} are satisfied. Additionally, suppose that there exists a positive-definite function $V(\rho)$ such that $V(\rho)=0$ if and only if $\rho=\bar{\boldsymbol\rho}$, and $V$ is continuous on $\mathcal{S}_N$ and twice continuously differentiable on the set $\mathcal S_N\setminus\{\bar{\boldsymbol\rho}\}$. Moreover, suppose that there exist positive constants $C$, $C_1$ and $C_2$ such that 
\begin{enumerate}
\item[(i)] $C_1 \, d_B(\rho,\bar{\boldsymbol\rho}) \leq V(\rho) \leq C_2 \, d_B(\rho,\bar{\boldsymbol\rho})$, $\forall\,\rho\in\mathcal{S}_N$, and 
\item[(ii)] $\limsup_{\rho\rightarrow\bar{\boldsymbol\rho}}\mathscr{L}V(\rho)/V(\rho)\leq-C$.
\end{enumerate}
Then, $\bar{\boldsymbol\rho}$ is almost surely exponentially stable for System~\eqref{SME} with sample Lyapunov exponent less or equal than $-C-K/2$, where $K:=\liminf_{\rho \rightarrow\bar{\boldsymbol\rho}}\sum^{m}_{i=1} g_i(\rho)^2$ with $g_i(\rho):=\sqrt{\eta_i}\Tr\left(\frac{\partial V(\rho)}{\partial \rho}\frac{G_i(\rho)}{V(\rho)}\right)$ for $i=1,\dots,m$.
\label{Thm a.s. exp stab}
\end{theorem}
\medskip

In the following, as applications of the previous theorem, we obtain explicit conditions on the feedback controller guaranteeing the exponential convergence and the corresponding estimates of the rate of convergence. Recalling that  $\bar{\boldsymbol\rho}=\mathbf{GHZ}_{\mathbf k}^{\boldsymbol\epsilon},$ where $\boldsymbol\epsilon\in\{+,-\}$ and $\mathbf k\in\{1,\dots,N/2\},$ we define $\Gamma_m:=\min_{i\in\{1,\dots,m\}}\{\eta_i M_i\}$, $c_{+}:=\frac{1}{\sqrt{m_z}}(l_{\mathbf{k}}-\max_{k\neq \mathbf{k}}\{l_k\})-1$ and $c_{-}:=\frac{1}{\sqrt{m_z}}(l_{\mathbf{k}}-\min_{k\neq \mathbf{k}}\{l_k\})+1.$
\begin{theorem}
Consider System~\eqref{SME} with $\rho_{0} \in \mathcal{S}_N$ and assume $\eta_1,\dots,\eta_m\in(0,1)$. Let $\bar{\boldsymbol \rho}=\mathbf{GHZ}_{\mathbf k}^{\boldsymbol\epsilon}$ be the target GHZ state. Suppose that either $c_+>0$ or $c_-<0,$ and the assumptions of Lemma~\ref{Passage lemma_2QC} are satisfied. Moreover, assume that 
\begin{equation}
\limsup_{\rho\rightarrow \bar{\boldsymbol{\rho}}}\Theta_u(\rho)/d_B(\rho,\bar{\boldsymbol{\rho}})^2=0,
\label{Relation u dB}
\end{equation}
where $\Theta_u(\rho)=u(\rho)\Tr\big(i[H_1,\rho]\bar{\boldsymbol{\rho}}\big)$. Then, $\bar{\boldsymbol{\rho}}$ is almost surely exponentially stable with sample Lyapunov exponent less than or equal to
\begin{itemize}
\item[(i)] $-2\bar C_+$ with $\bar C_+:=\Gamma_m(\min\{c_+,1\})^2$ if $c_+>0;$
\item[(ii)] $-2\bar C_-$ with $\bar C_-:=\Gamma_m(\max\{c_-,-1\})^2$ if $c_-<0.$
\end{itemize}
\label{Thm Special case}
\end{theorem}
\proof
We apply Theorem~\ref{Thm a.s. exp stab} with the Lyapunov function 
$$V_{\mathbf{k},\boldsymbol\epsilon}(\rho)=\sqrt{1-\Tr(\rho\bar{\boldsymbol{\rho}})}=\frac{\sqrt{2}}{2}\sqrt{(1-\Lambda_{\mathbf{k}})+(1-\boldsymbol\epsilon 2\mathbf{Re}\{\rho_{\mathbf{k},\bar{\mathbf{k}}}\})}.$$ Since $\frac{\sqrt{2}}{2}d_B(\rho,\bar{\boldsymbol{\rho}})\leq V_{\mathbf{k},\boldsymbol\epsilon}(\rho)\leq d_B(\rho,\bar{\boldsymbol{\rho}})$ for all $\rho \in\mathcal{S}_{N}$, then the condition \textit{(i)} of Theorem~\ref{Thm a.s. exp stab} is verified. Next, we show that the condition \textit{(ii)} of Theorem~\ref{Thm a.s. exp stab} holds true as well. 
The infinitesimal generator of the Lyapunov function is estimated  by
\begin{equation*}
\begin{split}
\mathscr{L}V_{\mathbf{k},\boldsymbol\epsilon}(\rho)\leq \,&\frac{u(\rho)\Tr\big(i[H_1,\rho]\bar{\boldsymbol{\rho}}\big)}{2V_{\mathbf{k},\boldsymbol\epsilon}(\rho)}\\
&-\frac{\Gamma_m \Tr(\rho\bar{\boldsymbol{\rho}})^2}{4V_{\mathbf{k},\boldsymbol\epsilon}(\rho)^3}\Big(\frac{1}{\sqrt{m_z}}|\sum^{m_z}_{i=1}P_i(\rho)|+|P_m(\rho)| \Big)^2,
\end{split}
\end{equation*}

We have $\sum^{m_z}_{i=1}P_i(\rho)=l_{\mathbf{k}}-\Tr(\mathbf{L}_z\rho)=l_{\mathbf{k}}-\sum^{N/2}_{n=1}l_n\Lambda_n$ and $P_m(\rho)=1-\boldsymbol\epsilon\sum^{N/2}_{n=1}2\mathbf{Re}\{\rho_{n,\bar{n}}\}.$
Since $\big|\sum_{n\neq \mathbf{k}}2\mathbf{Re}\{\rho_{n,\bar{n}}\}\big|\leq \sum_{n\neq \mathbf{k}}2\left|\mathbf{Re}\{\rho_{n,\bar{n}}\}\right| \leq \sum_{n\neq \mathbf{k}}\Lambda_n=1-\Lambda_{\mathbf{k}},$
we deduce the following results for two cases $c_{+}>0$ and $c_{-}<0$.

If $c_{+}:=\frac{1}{\sqrt{m_z}}(l_{\mathbf{k}}-\max_{k\neq \mathbf{k}}\{l_k\})-1>0$, then 
\begin{equation*}
\begin{split}
 &\frac{1}{\sqrt{m_z}}|\sum^{m_z}_{i=1}P_i(\rho)|+|P_m(\rho)|\geq\frac{1}{\sqrt{m_z}}\sum^{m_z}_{i=1}P_i(\rho)+|P_m(\rho)|\\
  &\geq c_+(1-\Lambda_{\mathbf{k}})+\big(1-\boldsymbol\epsilon2\mathbf{Re}\{\rho_{\mathbf{k},\bar{\mathbf{k}}}\}\big)\geq 2\min\{c_+,1\}V^2_{\mathbf{k},\boldsymbol\epsilon}(\rho),
\end{split}
\end{equation*}
then we have 
\begin{equation*}
\Big(\frac{1}{\sqrt{m_z}}|\sum^{m_z}_{i=1}P_i(\rho)|+|P_m(\rho)| \Big)^2 \geq 4\big(\min\{c_+,1\}\big)^2V^4_{\mathbf{k},\boldsymbol\epsilon}(\rho)
\end{equation*}
which implies that
\begin{equation*}
\begin{split}
\mathscr{L}V_{\mathbf{k},\boldsymbol\epsilon}(\rho)&\leq \frac{\Theta_u(\rho)}{2V_{\mathbf{k},\boldsymbol\epsilon}(\rho)}-\Gamma_m\big(\min\{c_+,1\}\big)^2\Tr(\rho\bar{\boldsymbol{\rho}})^2V_{\mathbf{k},\boldsymbol\epsilon}(\rho)\\
&\leq -\bar{C}_{+}\Big(\Tr(\rho\bar{\boldsymbol{\rho}})^2- \frac{\Theta_u(\rho)}{2\bar{C}_{+}V_{\mathbf{k},\boldsymbol\epsilon}(\rho)^2}\Big)V_{\mathbf{k},\boldsymbol\epsilon}(\rho),
\end{split}
\end{equation*}
where $\bar{C}_{+}:=\Gamma_m\big(\min\{c_+,1\}\big)^2$. By the equivalence between $V_{\mathbf{k},\boldsymbol\epsilon}(\rho)$ and $d_B(\rho,\bar{\boldsymbol{\rho}})$ and the condition~\eqref{Relation u dB}, we have $\limsup_{\rho\rightarrow \bar{\boldsymbol{\rho}}}\frac{\mathscr{L}V_{\mathbf{k},\boldsymbol\epsilon}(\rho)}{V_{\mathbf{k},\boldsymbol\epsilon}(\rho)}\leq -\bar{C}_{+}$. Moreover, we have $\sum^{m}_{i=1}g_i(\rho)^2\geq2\bar{C}_{+}\Tr(\rho\bar{\boldsymbol{\rho}})^2$, where $g_i(\rho)$ for $i=1,\dots,m$ are defined in Theorem~\ref{Thm a.s. exp stab}. This implies $K=2\bar{C}_{+}$.

Similarly, if $c_{-}:=\frac{1}{\sqrt{m_z}}(l_{\mathbf{k}}-\min_{k\neq \mathbf{k}}\{l_k\})+1<0$, then 
\begin{equation*}
\begin{split}
\mathscr{L}V_{\mathbf{k},\boldsymbol\epsilon}(\rho)\leq  -\bar{C}_{-}\Big(\Tr(\rho\bar{\boldsymbol{\rho}})^2- \frac{\Theta_u(\rho)}{2\bar{C}_{-}V_{\mathbf{k},\boldsymbol\epsilon}(\rho)^2}\Big)V_{\mathbf{k},\boldsymbol\epsilon}(\rho),
\end{split}
\end{equation*}
where $\bar{C}_{-}:=\Gamma_m(\max\{c_-,-1\})^2$. 
With the same reasoning provided above in $(i)$, we find $K=2\bar{C}_{-}$.

The proof is complete.\hfill$\square$
\medskip

By dropping the assumptions on the sign of  $c_+$ or $c_-$, we get the following result.
\begin{theorem}
Consider System~\eqref{SME} with $\rho_{0} \in \mathrm{int}(\mathcal{S}_N)$. Let $\bar{\boldsymbol\rho}=\mathbf{GHZ}_{\mathbf k}^{\boldsymbol\epsilon}$. Suppose that  the assumptions of Lemma~\ref{Passage lemma_2QC} are satisfied, and for all $\rho\in\mathcal{S}_N$, $|u(\rho)|\leq Cd_B(\rho,\bar{\boldsymbol\rho})^\zeta$ for some constants $\zeta>1$ and $C>0$. Then, $\bar{\boldsymbol{\rho}}$ is almost surely exponentially stable with sample Lyapunov exponent less than or equal to $-\bar C$ with $\bar C:=\min\{\Gamma_z \ell^2/2m_z,2\eta_m M_m\},$ where $\Gamma_z=\min_{i\in\{1,\dots,m_z\}}\{\eta_i M_i\}.$
\label{Thm General case}
\end{theorem}
\proof
Consider the candidate Lyapunov function $V_{\mathbf{k},\boldsymbol\epsilon}(\rho)=\sum^{N/2}_{n\neq \mathbf{k}}\sqrt{\Lambda_n}+\sqrt{1-\boldsymbol\epsilon\Tr(L_x\rho)}\geq 0.$
Due to Lemma~\ref{Lem:RankNonDec}, all diagonal elements of $\rho_t$ remain strictly positive for all $t\geq0$ almost surely. We remark that $V_{\mathbf{k},\boldsymbol\epsilon}(\rho)$ is $\mathcal C^2$ in $\mathrm{int}(\mathcal{S}_N),$ and we can show that $\frac{\sqrt{2}}{2}d_B(\rho,\bar{\boldsymbol{\rho}})\leq V_{\mathbf{k},\boldsymbol\epsilon}(\rho)\leq 2d_B(\rho,\bar{\boldsymbol{\rho}})$ for all $\rho \in\mathcal{S}_{N}$, which verifies the condition \textit{(i)} of Theorem~\ref{Thm a.s. exp stab}. 

In order to verify the condition \textit{(ii)} of Theorem~\ref{Thm a.s. exp stab}, we note that for any $n\neq\mathbf{k}$,
\begin{equation*}
\begin{split}
\mathscr{L}\sqrt{\Lambda_n}&=-\frac{u\Tr\big(i[H_1,\rho_t](\mathbf{GHZ}^{+}_n+\mathbf{GHZ}^{-}_n)\big)}{2\sqrt{\Lambda_n}}\\
&~~~~-\frac{\sqrt{\Lambda_n}}{2}\sum^{m_z}_{i=1}\eta_iM_iQ_n^{(i)}(\rho)^2\\
&~~~~ -\frac{\eta_mM_m}{2(\Lambda_n)^{3/2}}\big(2\mathbf{Re}\{\rho_{n,\bar{n}}\}-\Lambda_n\Tr(L_x\rho)\big)^2\\
&\leq -\frac{u(\rho)\Tr(i[H_1,\rho_t](\mathbf{GHZ}^{+}_n+\mathbf{GHZ}^{-}_n))}{2\sqrt{\Lambda_n}}\\
&~~~~-\frac{\Gamma_z\sqrt{\Lambda_n}}{2}\sum^{m_z}_{i=1}Q_n^{(i)}(\rho)^2,
\end{split}
\end{equation*}
where $Q_n^{(i)}(\rho):=l_n^{(i)}-\Tr(L_z^{(i)}\rho)$ for $i\in\{1,\dots,m_z\}.$ 

In some neighborhood of $\mathbf{GHZ}^{\epsilon}_{\mathbf{k}}$ for $n\neq\mathbf{k}$, we have
\begin{equation*}
\begin{split}
&m_z\sum^{m_z}_{i=1}Q_n^{(i)}(\rho)^2\geq(l_n-\Tr(\mathbf{L}_z\rho))^2\\
&\geq (|l_{n}-l_{\mathbf{k}}|-|l_{\mathbf{k}}-\Tr(\mathbf{L}_z\rho)|)^2\geq (\ell-|l_{\mathbf{k}}-\Tr(\mathbf{L}_z\rho)|)^2
\end{split}
\end{equation*}
Moreover, since $(\mathbf{Re}\{\rho_{m,n}\})^2+(\mathbf{Im}\{\rho_{m,n}\})^2\leq\rho_{m,m}\rho_{n,n}$ for $m,n\in\{1,\dots,N\}$ and the fact that $H_1$ and $\rho$ are Hermitian, after a straightforward calculation, we can show that, for all $\rho\in\mathcal{S}_N$ and for all $n\neq\mathbf{k}$, $|\Tr(i[H_1,\rho](\mathbf{GHZ}^{+}_n+\mathbf{GHZ}^{-}_n))|\leq c_n\sqrt{\Lambda_{n}}$, for some $c_n>0$. Then, we deduce that
\begin{equation*}
\begin{split}
&\mathscr{L}\left(\textstyle\sum^{N/2}_{n\neq \mathbf{k}}\sqrt{\Lambda_n}\right) \\
&\leq|u|\sum^{N/2}_{n\neq \mathbf{k}} c_n-\frac{\Gamma_z}{2m_z}(\ell-|l_{\mathbf{k}}-\Tr(\mathbf{L}_z\rho)|)^2\sum^{N/2}_{n\neq \mathbf{k}}\sqrt{\Lambda_n}.
\end{split}
\end{equation*}
We denote $V_x(\rho):=1-\Tr(L_x\rho)^2$ and $V_{\boldsymbol\epsilon}(\rho):=1+\boldsymbol\epsilon\Tr(L_x\rho)$ for $\boldsymbol\epsilon\in\{+,-\}.$
We find
\begin{equation*}
\begin{split}
\mathscr{L}&\sqrt{1-\boldsymbol\epsilon\Tr(L_x\rho)}\\
&=\frac{\boldsymbol\epsilon u\Tr(i[H_1,\rho]L_x)}{\sqrt{1-\boldsymbol\epsilon\Tr(L_x\rho)}}-\frac{ \eta_m M_m V_x(\rho)^2 }{2 \big( 1-\boldsymbol\epsilon\Tr(L_x\rho) \big)^{3/2}}\\
&~~~~-\frac{ \sum^{m_z}_{i=1}\eta_i \big(\Tr(L_x L_i \rho)-\Tr(L_x\rho )\Tr(L_i\rho) \big)^2 }{2 \big( 1-\boldsymbol\epsilon\Tr(L_x\rho) \big)^{3/2}}\\
&\leq 2\sqrt{2}\|H_1\sqrt{\rho}\|_{HS}|u| -\frac{\eta_m M_m V_x(\rho)^2}{2 \big( 1-\boldsymbol\epsilon\Tr(L_x\rho) \big)^{3/2}}\\
&\leq C|u|-\frac{\eta_m M_m V_{\boldsymbol\epsilon}(\rho)^2}{2}\sqrt{1 -\boldsymbol\epsilon\Tr(L_x\rho)},
\end{split}
\end{equation*}
where $C>0$ and the first term of the first inequality follows from
\begin{equation*}
\begin{split}
&|\Tr([H_1,\rho]L_x)|\\
&=|\Tr([H_1,\rho](L_x-\boldsymbol\epsilon\mathds{1}))|=|\Tr(H_1[\rho,L_x-\boldsymbol\epsilon\mathds{1}])|\\
 &\leq|\Tr(H_1\rho(L_x-\boldsymbol\epsilon\mathds{1}))|+|\Tr(H_1(L_x-\boldsymbol\epsilon\mathds{1})\rho)|\\
 &\leq 2\|H_1\sqrt{\rho}\|_{HS}\|(L_x-\boldsymbol\epsilon\mathds{1})\sqrt{\rho}\|_{HS}\\
 &=2\sqrt{2}\|H_1\sqrt{\rho}\|_{HS}\sqrt{1-\boldsymbol\epsilon\Tr(L_x\rho)},
\end{split}
\end{equation*}
where we have used Cauchy-Schwartz inequality and $\|\cdot\|_{HS}$ represents the Hilbert-Schmidt norm. 

By the assumption of the theorem on the feedback controller and the equivalence between the Lyapunov function and the Bures distance, we have $|u|\leq K V_{\mathbf{k},\boldsymbol\epsilon}(\rho)^{\zeta}$ for some $K>0$ and $\zeta>1$. Finally, we have the following estimation of the infinitesimal generator of $V_{\mathbf{k},\boldsymbol\epsilon}(\rho)$,
\begin{equation*}
\begin{split}
\mathscr{L}V_{\mathbf{k},\boldsymbol\epsilon}(\rho)&\!\leq\! |u|\Big(\sum^{N/2}_{n\neq \mathbf{k}} c_n\!+\!C\Big)\!-\!\frac{\eta_m M_m V_{\boldsymbol\epsilon}(\rho)^2}{2}\sqrt{1 -\boldsymbol\epsilon \Tr(L_x\rho)}\\
&~~~~-\!\frac{\Gamma_z}{2m_z}(\ell-|l_{\mathbf{k}}-\Tr(\mathbf{L}_z\rho)|)^2\sum^{N/2}_{n\neq \mathbf{k}}\sqrt{\Lambda_n}\\
&\leq  -\Big(\mathbf{C}^{\boldsymbol\epsilon}_\mathbf{k}(\rho)-K\Big( \sum^{N/2}_{n\neq \mathbf{k}} c_n+C \Big)V_{\mathbf{k},\boldsymbol\epsilon}(\rho)^{\zeta-1}  \Big)V_{\mathbf{k},\boldsymbol\epsilon}(\rho),
\end{split}
\end{equation*}
where 
\begin{equation*}
\begin{split}
\mathbf{C}^{\boldsymbol\epsilon}_\mathbf{k}(\rho):=\min\lbrace &\eta_m M_m V_{\boldsymbol\epsilon}(\rho)^2/2\,,\\
&~~\Gamma_z(\ell-|l_{\mathbf{k}}-\Tr(\mathbf{L}_z\rho)|)^2/2m_z\rbrace.
\end{split}
\end{equation*}
Note that, $\lim_{\rho\rightarrow\bar{\boldsymbol{\rho}}}V_{\boldsymbol\epsilon}(\rho)=2$, thus we have
\begin{equation*}
\limsup_{\rho\rightarrow\bar{\boldsymbol\rho}}\frac{\mathscr{L}V_{\mathbf{k},\boldsymbol\epsilon}(\rho)}{V_{\mathbf{k},\boldsymbol\epsilon}(\rho)}\leq-\min\{\Gamma_z \ell^2/2m_z\, ,\,2\eta_m M_m\}<0.
\end{equation*}
Therefore, the condition \textit{(ii)} of Theorem~\ref{Thm a.s. exp stab} is verified. The proof is complete.\hfill$\square$

\begin{remark}
Suppose $\rho_0\in\mathrm{int}(\mathcal{S}_N)$, and the assumptions on the feedback controller and the control Hamiltonian of Theorem~\ref{Thm Special case} are satisfied. By applying the arguments in the proof of Theorem~\ref{Thm General case}, we can show that $\bar{\boldsymbol{\rho}}$ is almost surely exponentially stable with sample Lyapunov exponent less than or equal to $-\min\{\Gamma_z \ell^2/2m_z\, ,\,2\eta_m M_m\}-\bar C_+$ if $c_+>0$ and $-\min\{\Gamma_z \ell^2/2m_z\, ,\,2\eta_m M_m\}-\bar C_-$ if $c_-<0$, where $\bar{C}_{\pm}$ are defined in Theorem~\ref{Thm Special case}.
\end{remark}
\subsection{Feedback controller design}
In this section, we give explicit forms  of feedback controllers satisfying the assumptions of~Theorem~\ref{Thm Special case} and Theorem~\ref{Thm General case}.
\begin{proposition}
Consider System~\eqref{SME} with $\rho_0\in\mathcal{S}_N$. Let $\bar{\boldsymbol\rho}=\mathbf{GHZ}_{\mathbf k}^{\boldsymbol\epsilon}$. Suppose that \textbf{(A1)} is satisfied, either $c_+>0$ or $c_-<0$, and the control Hamiltonian satisfies the assumptions of Lemma~\ref{Passage lemma_2QC}.
Define the feedback controller as 
\begin{equation}
u(\rho)=\alpha \big(1-\Tr(\rho\bar{\boldsymbol\rho})\big)^{\beta},
\label{FB:SPECIAL}
\end{equation}
where $\beta>1$ and $\alpha>0$. Then, $\bar{\boldsymbol{\rho}}$ is almost surely exponentially stable with sample Lyapunov exponent less than or equal to 
\begin{itemize}
\item[i)] $-2\bar C_+$ with $\bar C_+:=\Gamma_m(\min\{c_+,1\})^2$ if $c_+>0$, and 

\item[ii)] $-2\bar C_-$ with $\bar C_-:=\Gamma_m(\max\{c_-,-1\})^2$ if $c_-<0$.
\end{itemize}
\label{PROP:DESIGN_SPECIAL}
\end{proposition}

\begin{proposition}
Consider System~\eqref{SME} with $\rho_0\in\mathrm{int}(\mathcal{S}_N)$. Let $\bar{\boldsymbol\rho}=\mathbf{GHZ}_{\mathbf k}^{\boldsymbol\epsilon}$. Suppose that \textbf{(A1)} holds and the control Hamiltonian satisfies the assumptions of Lemma~\ref{Passage lemma_2QC}. Define the feedback controller as 
\begin{equation}
 u(\rho)=\alpha \big(l_{\mathbf k}-\Tr(\mathbf{L}_z\rho)\big)^{\beta}+\gamma \big(\boldsymbol\epsilon 1-\Tr(L_x\rho)\big)^{\delta},
\label{FB:GENERAL}
\end{equation}
where $\beta,\delta>1$ and $\alpha,\gamma>0$ sufficiently large. Then, $\bar{\boldsymbol{\rho}}$ is almost surely exponentially stable with sample Lyapunov exponent less than or equal to $-\bar C$ with $\bar C:=\min\{\Gamma_z \ell^2/2m_z,2\eta_m M_m\}$.
\label{PROP:DESIGN_GENERAL}
\end{proposition}

\section{Discussion on asymptotic stabilization in presence of only $z$-type measurements}
\label{SEC:1QC}
In this section, we denote the target state as $\bar{\boldsymbol\rho}=\mathbf{GHZ}_{\mathbf k}^{\boldsymbol\epsilon}$. We suppose only $m=m_z$ measurement operators $L_i=\sqrt{M_i}L^{(i)}_z$  are present and \textbf{(A1)} is satisfied. Due to the analysis in Section~\ref{Sec: Asymptotics without feedback}, System~\eqref{SME} with $u\equiv0$ converges to the union of the sets $\mathcal T_k=\{\rho\in\mathcal{S}_N|\,\Lambda_{k}=1\}$ for $k\in\{1,\dots,N/2\}$, each of which contains two GHZ states $\mathbf{GHZ}^{\pm}_{k}$. We assume the presence of  $n\geq2$ control Hamiltonians.

By employing similar arguments as those in the first  step of the proof of~\cite[Theorem 6.2]{liang2019exponential}, we obtain general Lyapunov-type conditions ensuring local stability in probability of the target state with only $z$-type measurements. Denote by $\mathcal{K}$ the family of all continuous non-decreasing functions $\mu:\mathbb{R}_{\geq 0}\rightarrow\mathbb{R}_{\geq0}$ such that $\mu(0)=0$ and $\mu(r)>0$ for all $r>0$.

\begin{lemma}
 Assume that there exists a twice continuously differentiable positive function $V(\rho)$ such that $V(\rho)=0$ if and only if $\rho=\bar{\boldsymbol\rho}$. Moreover, suppose that there exists $\mu\in\mathcal{K}$ such that 
\label{THM:AC_1OM}
\begin{enumerate}
\item[(i)] $V(\rho) \geq \mu\big(d_B(\rho,\bar{\boldsymbol\rho})\big)$ for all $\rho\in\mathcal{S}_N$, and 
\item[(ii)] $\mathscr{L}V(\rho)\leq 0$ for all $\rho \in B_r(\bar{\boldsymbol\rho})$ with $r>0$.
\end{enumerate}
Then, $\bar{\boldsymbol\rho}$ is locally stable in probability for System~\eqref{SME}.
\end{lemma}
\medskip

Inspired by~\cite[Theorem 5.1]{mirrahimi2007stabilizing}, we propose the following example satisfying the assumptions of the above lemma. We define 
a continuously differentiable function on $[0,1]$,
\begin{equation*}
f(x) := 
\begin{cases}
0,&\textrm{for } x\in[0,\epsilon_1);\\
\frac12\sin\left(\frac{\pi(2x-\epsilon_1-\epsilon_2)}{2(\epsilon_2-\epsilon_1)}\right)+\frac12,&\textrm{for } x\in[\epsilon_1,\epsilon_2);\\
1,&\textrm{for } x\in(\epsilon_2,1],\\
\end{cases}
\end{equation*}
where $0<\epsilon_1<\epsilon_2<1$.
\begin{example} 
Consider System~\eqref{SME} with $\rho_0\in\mathcal{S}_N$ and $n=2$. Consider the feedback laws 
\begin{equation}
\begin{split}
u_1(\rho) &= \gamma-\Tr(i[H_1,\rho]\bar{\boldsymbol\rho}),\\
u_2(\rho) &= f(\Tr(\rho\bar{\boldsymbol\rho}))(\gamma-\Tr(i[H_2,\rho]\bar{\boldsymbol\rho})).
\end{split}
\label{FB:1OM}
\end{equation}
We suppose $[H_1+H_2,\bar{\boldsymbol\rho}]=0$ and we consider the following Lyapunov function $V(\rho)=1-\Tr(\rho\bar{\boldsymbol\rho})$ whose infinitesimal generator, for all $\rho\in B_r(\bar{\boldsymbol\rho})$, is given by 
\begin{equation*}
\begin{split}
\mathscr{L}V(\rho)&=\sum^{2}_{k=1} u_k(\rho)\Tr(i[H_k,\rho]\bar{\boldsymbol\rho})\\
&=-\Tr(i[H_1,\rho]\bar{\boldsymbol\rho})^2-\Tr(i[H_2,\rho]\bar{\boldsymbol\rho})^2\leq0,
\end{split}
\end{equation*}
where $r>0$ is sufficiently small, and for the last equality we used $[H_1+H_2,\bar{\boldsymbol\rho}]=0$. Hence, Lemma~\ref{THM:AC_1OM} can be applied.
\label{THM:DESIGN_1OM}
\end{example}

\medskip

To ensure asymptotic stability of $\bar{\boldsymbol\rho}$, we need to prove the reachability of any neighborhood of the target state.  The proof of this property is not straightforward, since roughly speaking, compared to the situation in Section~\ref{sec:asgs}, the trajectories of the new system are more constrained due to the absence of random displacements generated by $L_x.$ In the following, we discuss some conditions which may be sufficient to conclude the reachability.

We suppose $H_0=\sum_{i=1}^{m}\alpha_i L_i$ with $\alpha_i\in\mathbb R.$ For $l\in\mathbb{Z}$ and $\xi\in\mathbb R^N,$ we define 
\begin{equation*}
\begin{split}
\mathbf{M}^z_{l,\xi}\!:=\![&\xi, H_1\xi, L^{(1)}_zH_1\xi, \dots, L^{(m_z)}_zH_1\xi ,\\
&\dots, H_1^l\xi, L^{(1)}_zH_1^l\xi,\dots, L^{(m_z)}_zH_1^l\xi].
\end{split}
\end{equation*}
Moreover, we  assume $l_{\mathbf{k}}>l_k$ or $l_{\mathbf{k}}<l_k$ for all $k\neq \mathbf{k}$ and $\eta_1,\dots,\eta_{m_z}\in(0,1)$. 

 The following condition ensures that $\bar{E}_{\mathbf{n}}\setminus\{\bar{\boldsymbol\rho}\}$ does not contain any equilibrium of System~\eqref{SME}. 
\begin{enumerate}
\item[(A)] $u\in\mathcal{C}^1(\mathcal{S}_N,\mathbb{R}^n)$, $u_1(\boldsymbol\rho)[H_1,{\boldsymbol\rho}]\neq0$  and $u_k\equiv 0$ for $k>1$ for all $\boldsymbol\rho\in \bar{E}_{\mathbf{n}}\setminus\{\bar{\boldsymbol\rho}\}.$
\end{enumerate}
Furthermore, we consider the following conditions.
\begin{enumerate}
\item[(B)] There exists at least one $i\in\{1,\dots,m_z\}$ such that $\nabla u_1\cdot G_i(\rho_0)\neq 0$, or $\nabla u_1\cdot \big( F_0(\rho_0)+\sum^{m_z}_{i=1}\widehat F_i(\rho_0)\big)\neq 0$ for any $\rho_0\in\{\rho\in\mathcal S_N| \, \Tr(\rho\bar{\boldsymbol\rho})=0, \,\,u_1=0\},$ and $u_k\equiv 0$ for $k>1$ if $\Tr(\rho\bar{\boldsymbol\rho})=0.$
\item[(C)] For all $k\in\{1,\dots,N/2\}$ and $\epsilon\in\{+,-\},$ there exists a finite $l\in\mathbb{Z}$ such that $\mathrm{rank}(\mathbf{M}_{l,\xi}^z)=N,$ where $\xi=\mathrm{ghz}_k^\epsilon.$
\end{enumerate}
As $H_0=\sum_{i=1}^{m}\alpha_i L_i,$ similar to the proof of Lemma~\ref{Passage lemma_2QC}, we can show that $\Tr(\rho_v(t)\bar{\boldsymbol\rho})>0$ for some $t>0$ arbitrarily small under the assumptions (B) and (C). 

Now consider the case $\rho_v(t)\notin\mathcal{T}_{\mathbf{k}}.$  
Since $l_{\mathbf{k}}>l_k$ or $l_{\mathbf{k}}<l_k$ for all $k\neq \mathbf{k},$ we deduce that $l_{\mathbf{k}}-\Tr(\mathbf{L}_z\rho_v(t))\neq 0$ and therefore $|P_n(\rho_v(t))|=|l^{(n)}_{\mathbf k}-\Tr(L^{(n)}_z\rho_v(t))|>0$ for at least one $n\in\{1,\dots,m_z\}$. Hence $\Tr(\dot{\rho}_v(t)\bar{\boldsymbol\rho})$ can be made arbitrarily large by appropriately choosing $v\in \mathcal V_m.$ 

In the set $\mathcal{T}_{\bf k}\setminus\{\bar{\boldsymbol\rho}\},$ the functions $P_i(\rho)$ vanish for all $i\in\{1,\dots,m\}.$ 
Hence in  this set, the dynamics of $\Tr(\rho_v\bar{\boldsymbol\rho})$ is simply given by the vector field $\sum^n_{k=1}u_k[H_k,\rho]$ and does not depend 
on $v.$ Therefore, we consider the following condition. 
 \begin{enumerate}
\item[(D)]
$\mathcal{T}_{\bf k}\setminus\{\bar{\boldsymbol\rho}\}$ does not contain complete integral curves of the vector field $\sum^n_{k=1}u_k[H_k,\rho]$. 
\end{enumerate}
Unfortunately, the above arguments are not sufficient to conclude that any arbitrary neigborhood of the target state can be reached in finite time. The main obstacle is that one needs to provide a control input $v$ ensuring that $\Tr(\rho_v\bar{\boldsymbol\rho})$ approaches any value arbitrarily close to one in finite time avoiding the possibility that $\rho_v$  runs into $\mathcal{T}_{\bf k}\setminus\bar{\boldsymbol\rho}.$ 
\section{Simulation of three-qubit systems}
\label{SEC:SIM}
In this section, we consider the case of a three-qubit system with both $z$-type and $x$-type measurements, and only $z$-type measurements. We take $L^{(1)}_z=\sigma_z\otimes\mathds{1}\otimes\sigma_z$ and $L^{(2)}_z=2\sigma_z\otimes\sigma_z\otimes\mathds{1}$, and we set $H_0=\omega \mathbf{L}_z=\omega(L^{(1)}_z+L^{(2)}_z)$ with $\omega=0.3$.

\begin{figure}[!t]
\centerline{\includegraphics[width=\columnwidth]{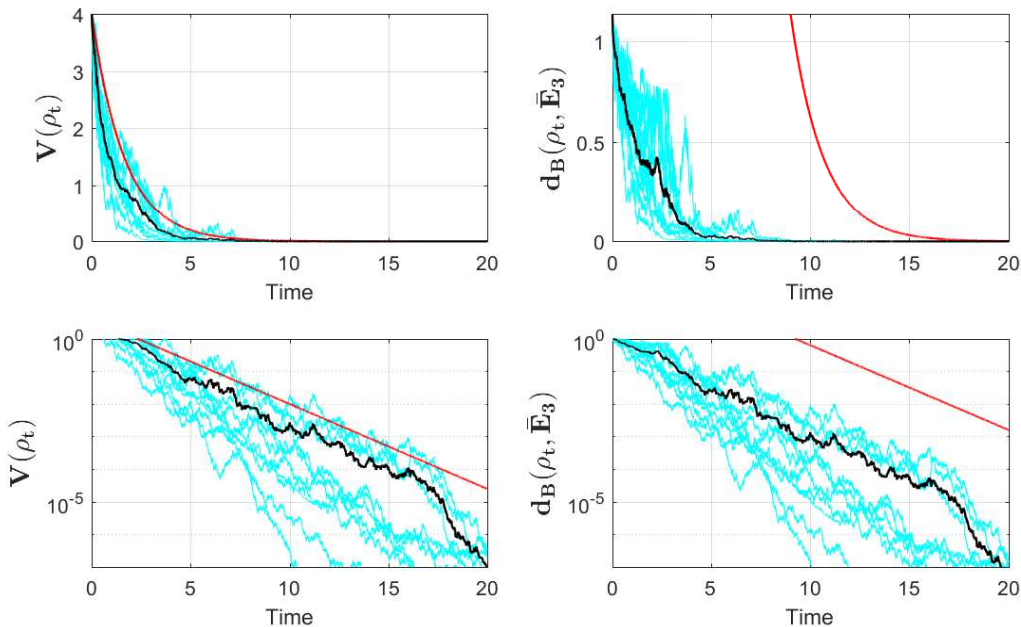}}
\caption{Quantum state reduction of a three-qubit system with $u\equiv 0$: the black curve represents the mean value of 10 arbitrary samples, the red curve represents the exponential reference with exponent $-0.3$. The figures at the bottom are the semi-log versions of the ones at the top. }
\label{SIM:QSR}
\end{figure}

\textit{ a) $z$-type and $x$-type measurements:} We assume $L_1=\sqrt{M_1}L^{(1)}_z$, $L_2=\sqrt{M_2}L^{(2)}_z$ and $L_3=\sqrt{M_3}L_x$ with $M_1=1.1$, $M_2=1$ and $M_3=0.9.$ We take $\eta_1=0.5$, $\eta_2=0.3$ and $\eta_3=0.4.$ We consider a Lyapunov function $V(\rho)$ of the form given in~\eqref{eq:lyapr}. The simulations with $u\equiv0$ starting from $\rho_0=\frac18\mathds{1}$ are shown in Figure~\ref{SIM:QSR}.
In particular, we observe that the expectation of the Lyapunov function $\mathbb{E}(V(\rho_t))$ is bounded by the exponential function $V(\rho_0)e^{-\bar{C}t}$ with $\bar{C}=\min\{\eta_1 M_1,\eta_2 M_2,2\eta_2 M_2\}$, and the expectation of the Bures distance $\mathbb{E}(d_B(\rho_t,\bar{E}_{3}))$ is always below the exponential function $C_2/C_1\,d_B(\rho_0,\bar{E}_3)e^{-\bar{C}t}$, with $C_1 = 1/8$ and $C_2 = 28$ in accordance with Theorem~\ref{QSR}.

We consider the control Hamiltonian $H_1 
= \big( \mathds{1}\otimes\mathds{1} + \mathds{1}\otimes\sigma_{x} + \sigma_{z}\otimes\sigma_{x} + \sigma_{z}\otimes\sigma_{y}\big)\otimes\sigma_{x}$.
By straightforward calculations, we have $[H_1,\rho]\neq 0$ for all $\rho\in\bar{E}_{3}$, and $\mathrm{rank}(\mathbf{M}_{3,\xi})=8$ for all $\xi=\mathrm{ghz}_k^\epsilon$ with $k\in\{1,\dots,4\}$ and $\epsilon=\pm$. If we choose $\bar{\boldsymbol\rho}=\mathbf{GHZ}^{\pm}_{1}$ we have $c_+=\sqrt{2}-1>0,$ and if we choose $\bar{\boldsymbol\rho}=\mathbf{GHZ}^{\pm}_{4}$ we have $c_-=1-\sqrt{2}<0$. 

\begin{figure}[!t]
\centerline{\includegraphics[width=\columnwidth]{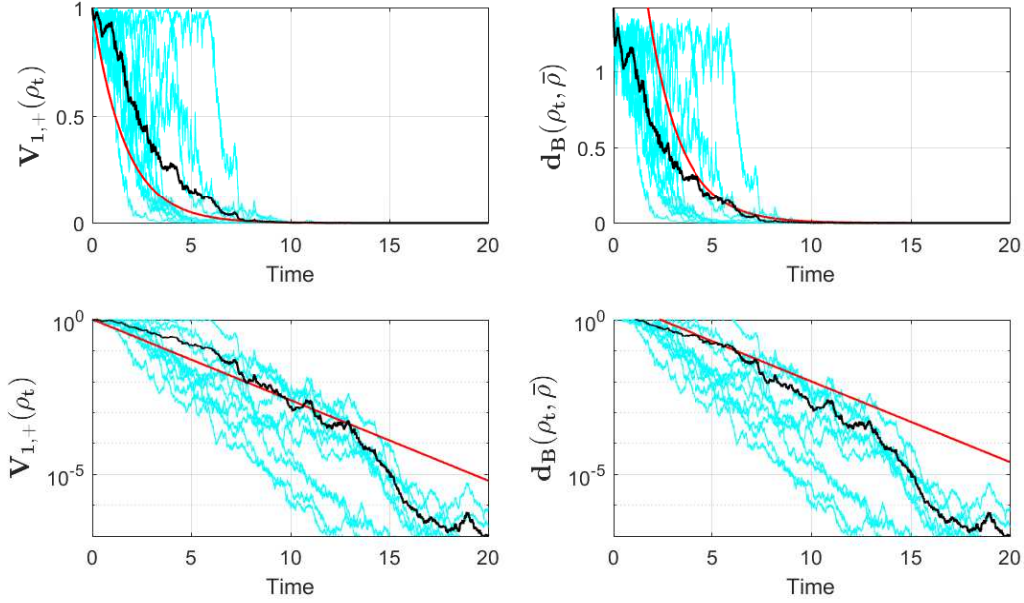}}
\caption{Exponential stabilization of a three-qubit system  towards $\mathbf{GHZ}^{+}_{1}$ with $u$ defined in~\eqref{FB:SPECIAL}: the black curve represents the mean value of 10 arbitrary samples, the red curve represents the exponential reference with exponent $\nu=-(9-6\sqrt{2})/5$. The figures at the bottom are the semi-log versions of the ones at the top. }
\label{SIM:FB1P}
\end{figure}

\begin{figure}[!t]
\centerline{\includegraphics[width=\columnwidth]{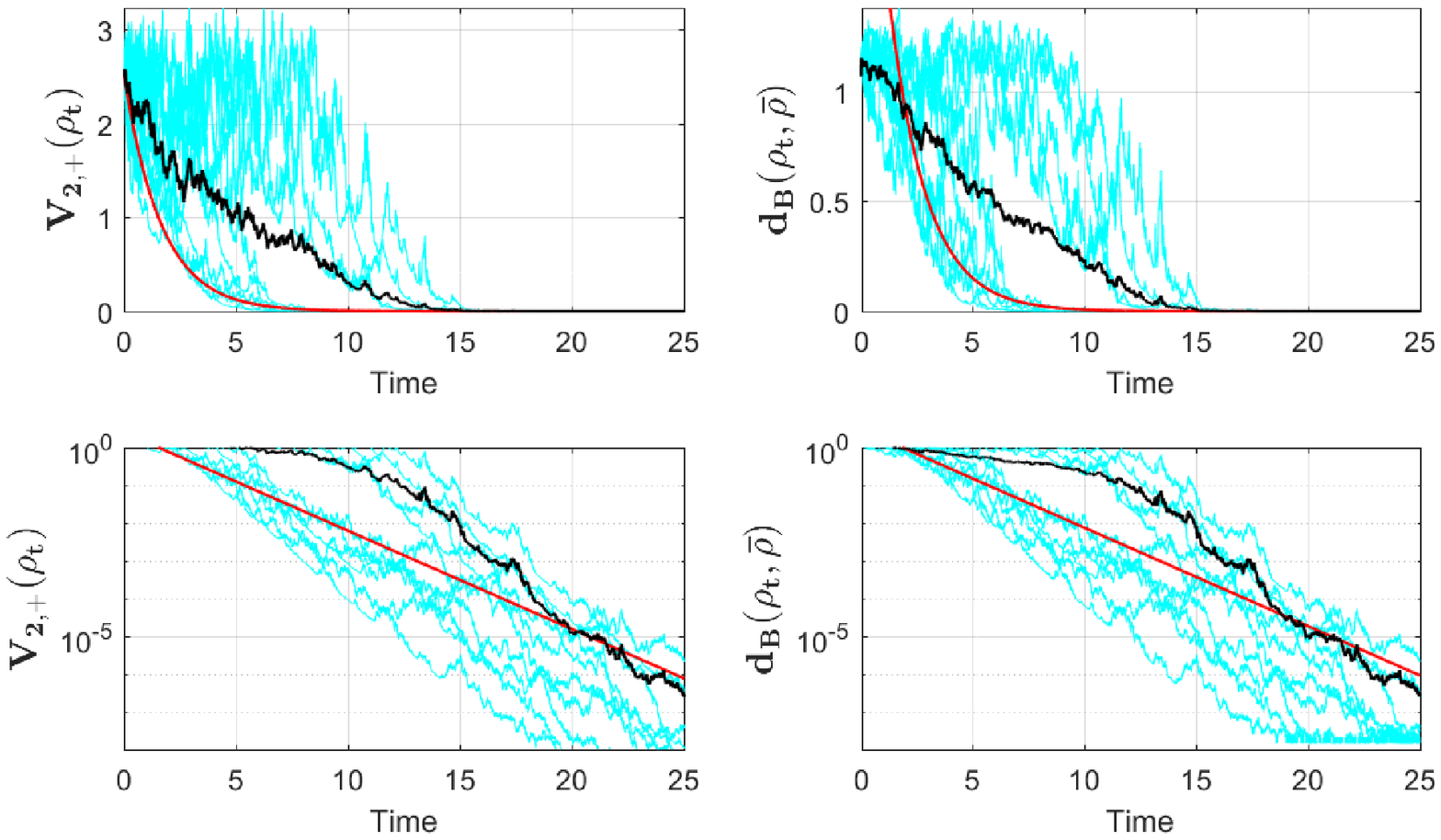}}
\caption{Exponential stabilization of a three-qubit system towards $\mathbf{GHZ}^{+}_{2}$ with $u$ defined in~\eqref{FB:GENERAL}: the black curve represents the mean value of 10 arbitrary samples, the red curve represents the exponential reference with exponent $\nu=-0.3$. The figures at the bottom are the semi-log versions of the ones at the top. }
\label{SIM:FB2P}
\end{figure}

In Fig.~\ref{SIM:FB1P}, we show the convergence of the system towards $\bar{\boldsymbol\rho}=\mathbf{GHZ}^{+}_{1}$ with a feedback controller~\eqref{FB:SPECIAL} with $\alpha=10$ and $\beta=7,$ and starting at $\mathbf{GHZ}^{-}_{4}\in\partial \mathcal{S}_N$. Then, the convergence of the system towards $\bar{\boldsymbol\rho}=\mathbf{GHZ}^{+}_{2}$ with a feedback controller~\eqref{FB:GENERAL} with $\alpha=\gamma=1$ and $\beta=\delta=5,$ and starting at $\frac18\mathds{1}\in\mathrm{int}(\mathcal{S}_N)$  is shown in~Fig.~\ref{SIM:FB2P}. We observe that the behaviors of the Lyapunov function and the Bures distance along the sample trajectories are consistent with the convergence rate estimates $\nu=-(6-4\sqrt{2})\min\{\eta_1 M_1,\eta_2 M_2,\eta_3M_3\}$  given in Theorem~\ref{Thm Special case} (Fig.~\ref{SIM:FB1P}), and $\nu=-\min\{\eta_1 M_1, \eta_2 M_2, 2\eta_3 M_3\}$  given in  Theorem~\ref{Thm General case} (Fig.~\ref{SIM:FB2P}). In the figures, the red curves represent the exponential reference with the exponent $\nu$, the black curves describe the mean values of the Lyapunov functions and the Bures distances of ten samples. On the figures, in particular in the semi-log versions, we can see that the cyan (sample trajectories) and the red curves have similar asymptotic behaviors.

\begin{figure}[!t]
\centerline{\includegraphics[width=\columnwidth]{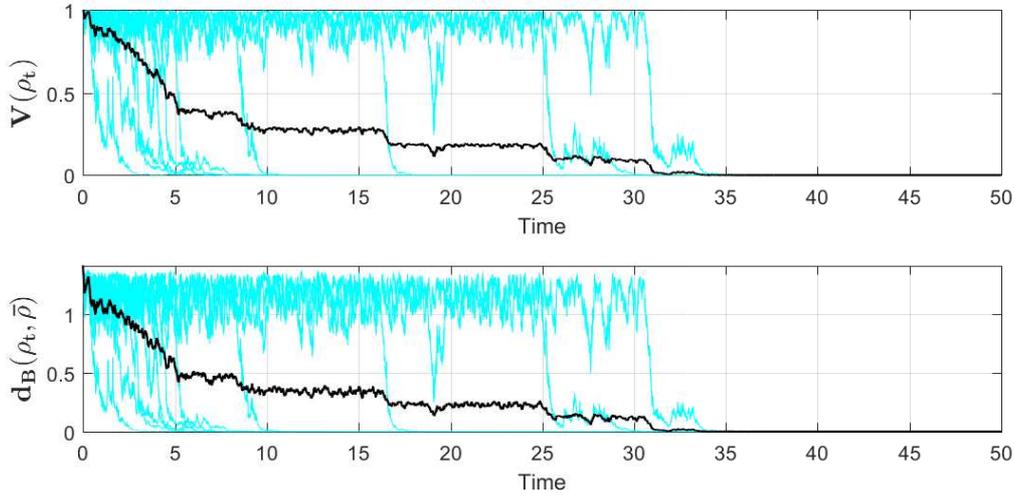}}
\caption{Asymptotic stabilization of a three-qubit system towards $\mathbf{GHZ}^{+}_{1}$ with $u$ defined in~\eqref{FB:1OM}: the black curve represents the mean value of 10 arbitrary samples.}
\label{SIM_FB1P_1QC}
\end{figure}

\textit{b) $z$-type measurements:}
In this case, we have only the measurement operators $L_1=\sqrt{M_1}L^{(1)}_z$ and $L_2=\sqrt{M_2}L^{(2)}_z$ with $M_1=1.1$ and $M_2=1$. We set  $\eta_1=0.5$, $\eta_2=0.3$ and $\mathbf{GHZ}^{+}_{1}$ as the target state, and define the following control Hamiltonians 
\begin{equation*}
\begin{split}
H_1 =& \big( \mathds{1}\otimes\mathds{1} + \mathds{1}\otimes\sigma_{x} + \sigma_{z}\otimes\sigma_{x} + \sigma_{z}\otimes\sigma_{y}\big)\otimes\sigma_{x},\\
H_2 =& -\sigma_{x}\otimes\sigma_{x}\otimes\mathds{1} - \mathds{1}\otimes\sigma_{x}\otimes\sigma_{x} \\
&- \sigma_{z}\otimes\sigma_{x}\otimes\sigma_{x} - \sigma_{y}\otimes\sigma_{z}\otimes\mathds{1}.
\end{split}
\end{equation*}
By straightforward calculations, we have $[H_1,\rho]\neq 0$ for all $\rho\in\bar{E}_{3}$, and we find $\mathrm{rank}(\mathbf{M}^z_{4,\xi})=8$ for all $\xi=\mathrm{ghz}_k^\epsilon$ with $k\in\{1,\dots,4\}$ and $\epsilon=\pm$. We can check  that the conditions (A)-(D) of Section~\ref{SEC:1QC} hold true as well as the assumption $[H_1+H_2,\bar{\boldsymbol\rho}]=0$ in Example~\ref{THM:DESIGN_1OM}. The convergence of the system towards $\mathbf{GHZ}^{+}_{1}$ with the feedback controller~\eqref{FB:1OM} with $\gamma=5,$ starting at $\mathbf{GHZ}^{-}_{4}\in\partial\mathcal{S}_N$  is shown in~Fig.~\ref{SIM_FB1P_1QC}.

\section{Conclusion and perspectives}
In this paper, we have studied the exponential and asymptotic stabilization of GHZ states. In presence of both $z$-type and $x$-type measurements and of one control Hamiltonian, general conditions on the feedback controller and control Hamiltonian guaranteeing exponential stabilization were provided, with estimations of the rate of convergence.  Moreover, parametrized continuous feedback controllers satisfying such conditions were proposed. For the case of only $z$-type measurements and multiple control Hamiltonians, we discussed the possibility of asymptotically stabilizing the system towards the target GHZ state. 

Further research lines will address the possibility of extending our results in presence of delays, and will focus on the robustness of our feedback scheme for multi-qubit systems~\cite{liang2020robustness_two,liang2020robustness_N}.
\section{Acknowledgements}
This work is supported by the Agence Nationale de la Recherche projects Q-COAST ANR-19-CE48-0003 and QUACO ANR-17-CE40-0007.

\bibliographystyle{alpha}
\bibliography{Ref_GHZ}

\appendix
\section{Proof of Inequality~\eqref{Eq:LV_QSR}}
\label{App:ComputationQSR}
Due to the cyclic property of the trace and the commutativity between $H_0$, all measurement operators $L_k$ and all GHZ states, we have $\sum^{m}_{k=0}\mathrm{Tr}\big((\mathbf{GHZ}^{+}_n+\mathbf{GHZ}^{-}_n)F_k(\rho)\big)=0$ when $u\equiv0$.
Then for any $n\in\{1,\dots,N/2\}$, the dynamics of $\Lambda_n(\rho_t)=\mathrm{Tr}((\mathbf{GHZ}^{+}_n+\mathbf{GHZ}^{-}_n)\rho_t)$ when $u\equiv0$ are given by
\begin{equation*}
\begin{split}
d\Lambda_n(\rho_t)&=\mathrm{Tr}((\mathbf{GHZ}^{+}_n+\mathbf{GHZ}^{-}_n)d\rho_t)\\
&=\sum^{m}_{k=1}\sqrt{\eta_k}\mathrm{Tr}\big((\mathbf{GHZ}^{+}_n+\mathbf{GHZ}^{-}_n)G_k(\rho_t) \big)dW_k(t)\\
&=2\Lambda_n(\rho_t)\sum^{m}_{k=1}\sqrt{\eta_k}\mathsf{P}^{(k)}_n(\rho)dW_k(t)\\
\end{split}
\end{equation*}
with $\mathsf{P}^{(k)}_n(\rho):=\mathfrak{l}^{(k)}_n-\mathrm{Tr}(L_k\rho)$ where $L_k(\mathbf{GHZ}^{+}_n+\mathbf{GHZ}^{-}_n)=\mathfrak{l}^{(k)}_n(\mathbf{GHZ}^{+}_n+\mathbf{GHZ}^{-}_n)$ and $\mathfrak{l}^{(k)}_n:=\sqrt{M_k}l^{(k)}_n$ for $k\in\{1,\dots,m_z\}$ and $\mathfrak{l}^{(m)}_n=0$.
By It\^o product rule, we have
\begin{equation*}
\begin{split}
&d\big(\Lambda_i(\rho_t)\Lambda_j(\rho_t) \big)\\
&=\Lambda_j(\rho_t)d\Lambda_i(\rho_t)+\Lambda_i(\rho_t)d\Lambda_j(\rho_t)+d\Lambda_i(\rho_t)d\Lambda_j(\rho_t)\\
&=2\Lambda_i(\rho_t)\Lambda_j(\rho_t)\sum^{m}_{k=1}\Big(\sqrt{\eta_k}\big(\mathsf{P}^{(k)}_i(\rho_t)+\mathsf{P}^{(k)}_j(\rho_t) \big)dW_k(t)\\
&~~~~~~~~~~~~~~~~~~~~~~~~~~~~+2\eta_k\mathsf{P}^{(k)}_i(\rho_t)\mathsf{P}^{(k)}_j(\rho_t)dt\Big).
\end{split}
\end{equation*}
Due to the invariance of the set $\mathcal{S}_{I}$ defined in the proof of Theorem~\ref{QSR}, we apply the It\^o formula to $\sqrt{\Lambda_j(\rho_t)\Lambda_i(\rho_t)}$ with $i\neq j$ and obtain the following infinitesimal generator
\begin{equation*}
\begin{split}
\mathscr{L}\sqrt{\Lambda_j(\rho)\Lambda_i(\rho)}&=-\sqrt{\Lambda_j(\rho)d\Lambda_i(\rho)}\sum^{m}_{k=1}\frac{\eta_k}{2}\big( \mathsf{P}^{(k)}_i(\rho)-\mathsf{P}^{(k)}_j(\rho) \big)^2\\
&=-\sqrt{\Lambda_j(\rho)\Lambda_i(\rho)}\sum^{m_z}_{k=1}\frac{\eta_kM_k}{2}\big( l^{(k)}_i-l^{(k)}_j \big)^2\\
&\leq -\frac{\Gamma_z}{2}\sqrt{\Lambda_j(\rho)\Lambda_i(\rho)}\sum^{m_z}_{k=1}\big( l^{(k)}_i-l^{(k)}_j \big)^2\\
&\leq -\frac{\Gamma_z \ell^2}{2m_z}\sqrt{\Lambda_j(\rho)\Lambda_i(\rho)}
\end{split}
\end{equation*}
where $\Gamma_z:=\min_{k\in\{1,\dots,m_z\}}\{\eta_k M_k\}$, and we used $\sum^{m_z}_{k=1}x_k^2\geq \frac{1}{m_z}\big(\sum^{m_z}_{k=1}x_k\big)^2$ for the last inequality. It implies 
\begin{equation*}
\mathscr{L}\sum_{i\neq j}\sqrt{\Lambda_j(\rho)\Lambda_i(\rho)}\leq -\frac{\Gamma_z \ell^2}{2m_z}\sum_{i\neq j}\sqrt{\Lambda_j(\rho)\Lambda_i(\rho)}.
\end{equation*}

The dynamics of $\mathrm{Tr}(L_x\rho_t)$ when $u\equiv 0$ is given by
\begin{equation*}
\begin{split}
d\mathrm{Tr}(L_x\rho_t)=&2\sum^{m_z}_{k=1}\sqrt{\eta_k}\Delta_k(\rho_t)dW_k(t)\\&+2\sqrt{\eta_mM_m}V_x(\rho_t)dW_m(t)
\end{split}
\end{equation*}
where $\Delta_k(\rho):=\mathrm{Tr}(L_kL_x\rho)-\mathrm{Tr}(L_k\rho)\mathrm{Tr}(L_x\rho)$ and $V_x(\rho)=1-\mathrm{Tr}(L_x\rho)^2\geq 0$.
Since the invariant of sets $\{\rho\in\mathcal{S}_N|\,V_x(\rho)=0\}$ and $\{\rho\in\mathcal{S}_N|\,V_x(\rho)>0\}$, we can apply It\^o formula to $\sqrt{V_x(\rho_t)}$ and obtain the following infinitesimal generator
\begin{equation*}
\begin{split}
\mathscr{L}\sqrt{V_x(\rho)}&=-2\eta_mM_m\sqrt{V_x(\rho)}-4\sum^{m_z}_{k=1}\frac{\eta_k\Delta_k(\rho)^2}{V_x(\rho)^{3/2}}\\
&\leq-2\eta_mM_m\sqrt{V_x(\rho)}.
\end{split}
\end{equation*}
Therefore, for $V(\rho)=\sum_{k\neq h}\sqrt{\Lambda_k(\rho)\Lambda_h(\rho)}+\sqrt{V_x(\rho)}$, we have
\begin{align*}
\mathscr{L}V(\rho)&\leq -\frac{\Gamma_z\ell^2}{2m_z} \sum_{k\neq h}\sqrt{\Lambda_k(\rho)\Lambda_h(\rho)}\\
&~~~~~~~~~~-2\eta_mM_m\sqrt{V_x(\rho)}\\
&\leq -\bar{C}V(\rho).
\end{align*}
where $\bar{C}:=\min\{\Gamma_z \ell^2/2m_z,2\eta_mM_m\}$.

\end{document}